\documentclass[AMA,STIX1COL]{WileyNJD-v2}

\newcommand{\sigmaF}{{\bm{\sigma_f}}}
\newcommand{\sigmaS}{{\bm{\sigma_s}}}
\newcommand{\vm}{\bm{v_m}}
\newcommand{\uu}{\bm{u}}
\newcommand{\vv}{\bm{v}}

\definecolor{bluefluid}{RGB}{179, 224, 255}
\definecolor{redsolid}{RGB}{230,10,10}

\usepackage{comment}
\usepackage{tikz}
\usetikzlibrary{decorations.pathreplacing}
\usepackage{makecell}
\usepackage{bm}
\usepackage{algorithm}
\usepackage{xcolor}
\usepackage{hyperref}
\usepackage{algpseudocode}
\usepackage{caption}
\usepackage{tabularx}
\usepackage{booktabs}
\articletype{Article Type}%

\begin{document}

\title{Proposal for Numerical Benchmarking of Fluid–Structure Interaction in Cerebral Aneurysms}

\author[1]{Aurèle Goetz}
\author[1]{Pablo Jeken Rico}
\author[2]{Yves Chau}
\author[2]{Jacques Sédat}
\author[1]{Aurélien Larcher}
\author[1]{Elie Hachem}

\authormark{Goetz \textsc{et al}}

\address[1]{\orgdiv{Computing and Fluids Research Group}, \orgname{CEMEF, Mines Paris PSL}, \orgaddress{\state{06904 Sophia Antipolis}, \country{France}}}

\address[2]{\orgdiv{Department of Neuro-Interventional and Vascular Interventional}, \orgname{University Hospital of Nice}, \orgaddress{\state{06000 Nice}, \country{France}}}

\corres{Address: 1 Rue Claude Daunesse, 06904 Sophia Antipolis, France.\\ *\email{aurele.goetz@minesparis.psl.eu}}

\fundingInfo{European Union (ERC, CURE, 101045042). Views and opinions expressed are however those of the author(s) only and do not necessarily reflect those of the European Union or the European Research Council. Neither the European Union nor the granting authority can be held responsible for them.}
 
\abstract[Summary]{Computational fluid dynamics is intensively used to deepen the understanding of aneurysm growth and rupture in the attempt to support physicians during therapy planning.
Numerous studies have assumed fully-rigid vessel walls in their simulations, whose sole hemodynamics may fail to provide a satisfactory criterion for rupture risk assessment.
Moreover, direct in-vivo observations of intracranial aneurysm pulsation have been recently reported, encouraging the development of fluid-structure interaction for their modelling and for new assessments. In this work, we describe a new fluid-structure interaction benchmark setting for the careful evaluation of different aneurysm shapes. The conﬁgurations consist of three real aneurysm domes positioned on a toroidal channel. All geometric features, meshing characteristics, ﬂow quantities, comparisons with rigid wall model and corresponding plots are provided. The results emphasize the alteration of flow patterns and hemodynamic descriptors between the different cases and in particular with the rigid-wall model, thereby underlining the importance of fluid-structure interaction modelling.} 


\maketitle

\section{Introduction}

Intracranial aneurysms (IAs) are pathological dilations of blood vessels that bear the risk of rupture and subsequent subarachnoidal haemorrhage, which is associated with high mortality and morbidity rates \cite{Ing+00}.
According to prevalence studies, around 3\% of the world population hosts at least one of these aneurysms \cite{Vla+11}. 
When identified, they raise the question of a potential clinical intervention, which also entails a non-negligible threat \cite{Kot+12, Nag+12}.
As a result, there is a need for risk-evaluation tools in order to comprehensively assess the stability of IAs.
So far, clinical decisions have been mostly based on the size, shape and location of the bulge.
However, it has been shown that the risk evaluation accuracy is limited when using these parameters only \cite{Nie+18}.
That is the reason why research effort is currently made to model and simulate patient-specific inner haemodynamics of IAs through Computational Fluid Dynamics (CFD) in order to aid physicians in decision-making.

Computational models of aneurysm biomechanics indeed hold great promise for risk stratification, as haemodynamic features reveal key correlations with future aneurysm growth \cite{Ceb+14, Men+14}.
Performing numerical simulations for large numbers of aneurysm cases has raised multiple challenges in the scientific community ranging from the efficient and systematic generation of adapted computational meshes to the solving of coupled systems of equations complemented with complex rheology models and tailored boundary conditions \cite{Jan+15}.
Most of these challenges aim at enriching the simulation fidelity towards real patient-specific predictive modelling.
Among them, moving from rigid arterial wall modelling to adequate Fluid-Structure Interaction (FSI) simulations stands as a key research goal.
Pure CFD modeling based on rigid wall assumption has been shown to overestimate Wall Shear Stresses (WSS) \cite{Tor+09, Baz+10}, which in turn cast doubt in its ability to provide satisfactory criteria for rupture risk assessment.
Fidelity can be refined by simulating the vascular flow in conjunction with vessel wall deformation via relevant coupled FSI modeling.
The early research effort in that field has been carried out by Torii et al. \cite{Tor+08, Tor+09, Tor+10}.
Authors have compared the results of fully-coupled FSI simulations using elastic and hyperelastic neo-Hookean wall behaviour \cite{Tor+08}.
They investigated later three aneurysms through elastic FSI simulations, comparing the obtained results to a fully-rigid configuration \cite{Tor+09}.
It has been suggested that the need for FSI modelling is geometry-dependent, with WSS in areas of flow impingement being overestimated in the rigid configuration.
Going a step further, a few studies reported the specific interest of FSI simulations, when the wall thickness is modelled accurately \cite{Baz+10, VoS+16}.
Voß et al. imaged the geometry of a single aneurysm dome using micro-CT after the tissue was resected in surgical clipping \cite{VoS+16}.
Subsequent FSI simulations of the acquired geometry compared the results obtained with a uniform-thickness configuration ($0.3\,mm$) and with the specific thickness distribution measured in micro-CT, revealing peak local stress variations of around 50\%.
A few other research teams have proposed similar studies \cite{Baz+10, Val+13} but always investigating different cohorts of aneurysm cases composed of a few specimens only.
This, along with different modelling assumptions, undermines comparisons between them, thereby limiting reproducibility of the reported results.
At the same time, it is important to recall that FSI in intracranial aneurysms is a complex multi-parametric problem, which would require large cohorts of investigated cases in order to draw robust conclusions.
However, simulating fluid-structure coupled physics involves a certain algorithmic complexity and is computationally costly.
In addition, clinical routine imaging techniques did not allow until recently the visualization of wall movement in the brain, thereby not encouraging the development of arterial compliance modelling for intracranial aneurysms.
As a result, no general consensus has been reached yet on the relevance of FSI modelling in the context of IA risk assessment.

Over the past few years, IA pulsations have been directly observed through clinical imaging \cite{Van+15, Hay+13, Zho+22, Sta+21}. This provides new perspectives to the FSI modelling of brain aneurysms.
Indeed, in-vivo data acquisition has substantially benefited the modelling of aortic aneurysms, with the prescription of patient-specific wall thicknesses \cite{Fin+13} and even local tissue stiffness estimation through 4D-flow analysis.
However, this has hardly been conceivable in the brain, as most aneurysms have wall thicknesses ranging between $30\,\mu m$ and $400\,\mu m$ \cite{Isa+08, SO78}, hence falling under common medical imaging resolutions.
Mostly supported by the development of very precise ECG-gated 4D-CTA \cite{Hay+13, Zho+22}, the situation is progressively changing and future research will surely benefit from additional in-vivo data to feed FSI models.
So far, aneurysm pulsation has been mostly expressed in terms of overall bulge volume variation over a cardiac cycle, with peak reported values of 20\% \cite{Sta+21}.
Even though these measurements suffer from large uncertainty, especially for small bulges \cite{Kleinloog2018, Illies2016}, this new insight motivates the development of FSI modelling of IAs beyond the scarce existing literature.
Furthermore, publications have already reported that IAs demonstrate very different mechanical properties and thicknesses \cite{SO78, Rob+15}, due to several biological phenomena linked with their formation and growth \cite{Men+14}, inducing potential local weaknesses in pathological tissue.
If future medical equipment allows the in-vivo localization of these weaker spots as done post-mortem in \cite{VoS+16}, FSI models will surely contribute to build precise rupture risk assessment tools.
On top of that, even if measurement data is still lacking, assessing the sensitivity of several physical parameters in the context of FSI will give crucial insights for the future of IA modelling.

In this work, we describe a new benchmark setting using idealized geometry for the careful evaluation of different aneurysm dome shapes in interaction with the blood flow dynamics.
Although the analysis of patient-specific geometries remains an ultimate goal, it seems that idealized IA geometries are still missing for studying FSI-related phenomena in a more controlled manner.
Idealized geometries of sidewall aneurysms have been investigated almost exclusively through rigid-wall CFD simulations \cite{Mer+10, Has+05, Ram+13}, showing the sensitivity of haemodynamics towards various geometrical parameters.
They proved to be very useful in particular for studying the impact of certain modelling assumptions and boundary conditions \cite{Ram+13}, or highlighting specific trends that could be reproduced.
In this work, we propose a novel FSI modelling based on the Variational Multiscale Method (VMS) for both the fluid and solid solvers.
It is employed to investigate the proposed benchmark setting along with three real aneurysm domes.
The introduced benchmark geometry is extremely versatile, as bulge shapes can be substituted while keeping the general case settings unchanged.
The simplified problem offers a better environment to draw conclusions from a smaller manifold of explored configurations.
All geometric features, meshes, flow quantities, comparisons with rigid wall model and corresponding plots are provided.
We cared that the test cases are easy to reproduce but help draw the necessary conclusions on the system’s sensitivity to physical and geometrical parameters.
The design of the proposed benchmark is described in the following section, along with the employed methods for solving the coupled physics.
Then, after validation of the method on the well-established pressure wave benchmark proposed in \cite{formaggia}, we explore the possibilities of the proposed geometry and shed light on bulge shapes that might consequently benefit from the modelling of compliant arterial tissue.

\clearpage
\section{Materials and methods}
\subsection{Design of the proposed FSI case}
\subsubsection{A simple but versatile geometry}

\begin{figure}[ht!]
    \centering
    
    \begin{tikzpicture}[scale=1]
    \begin{scope}
        \clip (-6,0) rectangle (6,6);
        \draw[fill=redsolid] circle (5.9cm);
        \draw[fill=bluefluid, ultra thick] circle (5.7cm);
        \draw[densely dotted] circle (4.3cm);
        \draw[fill=redsolid, ultra thick] circle (2.9cm);
    \end{scope}
    \begin{scope}
        \clip (-6,-0.005) rectangle (6,6);
        \draw[fill=white] circle (2.7cm);
    \end{scope}
    \draw [](-5.9,0) -- (-2.7,0);
    \draw [](5.9,0) -- (2.7,0);
    \begin{scope}
        \clip (-6,5.573) rectangle (6,10);
        \draw[fill=redsolid] (0,7) circle (2.4cm);
    \end{scope}
    \begin{scope}
        \clip (-6,5.462) rectangle (6,10);
        \draw[fill=bluefluid, ultra thick] (0,7) circle (2.2cm);
    \end{scope}

    \draw [<->](-4.3,-0.1) -- (-0,-0.1);
    \node at (-2.15,-0.4) {$R_{torus}$};
    \draw [<->](-4.3,-0.1) -- (-5.7,-0.1);
    \node at (-5,-0.4) {$r_{torus}$};
    \draw [<->](-5.9,-0.1) -- (-5.7,-0.1);
    \node at (-5.8,-0.4) {$\epsilon$};
    \draw [<->](0,7) -- (0,0);
    \node at (1,3.5) {$H_{aneurysm}$};
    \draw [<->](0,7) -- (2.2/1.41,7+2.2/1.41);
    \node at (1.2,7) {$r_{aneurysm}$};

    \node at (-3,3) {$\Omega_f$};
    \draw [-stealth](3.8,5.2) -- (2.88,5);
    \node at (4.2,5.2) {$\Omega_s$};

    \draw [-stealth](-4.1,0.2) -- (-4.2,0);
    \node at (-3.7,0.4) {$\Gamma_{f,in}$};
    \draw [-stealth](4.1,0.2) -- (4.2,0);
    \node at (3.7,0.4) {$\Gamma_{f,out}$};
    \draw [-stealth](-2.6,4.5) -- (-2.9,4.88);
    \node at (-2.1,4.3) {$\Gamma_{FSI}$};
    \draw [-stealth](-3.8,5.3) -- (-3,5.08);
    \node at (-4.4,5.3) {$\Gamma_{s,ext}$};

    \draw [-stealth, thick](0.8,0.5) -- (1.5,0.5);
    \draw [-stealth, thick](0.8,0.5) -- (0.8,1.2);
    \draw [fill=white, thick](0.8,0.5) circle (0.15cm);
    \draw[fill=black, ultra thick] (0.8,0.5) circle (0.03cm);
    \node at (0.8,1.4) {$\bm{e}_y$};
    \node at (1.8,0.5) {$\bm{e}_x$};
    \node at (0.6,0.2) {$\bm{e}_z$};

    \node at (4.5,7.3) {
        
        \begin{tabular}{ccc}
            $R_{torus}$  & = & 4.3\\
            $r_{torus}$ & = & 1.4\\
            $\epsilon$ & = & 0.2\\
            $H_{aneurysm}$ & = & 7\\
            $r_{aneurysm}$ & = & 2.2\\
        \end{tabular}
    };
    
    \end{tikzpicture}
    \caption{Schematic illustration of the proposed benchmark (dimensions are given in mm).}
    \label{fig:AnXplore_reference_definition}
\end{figure}
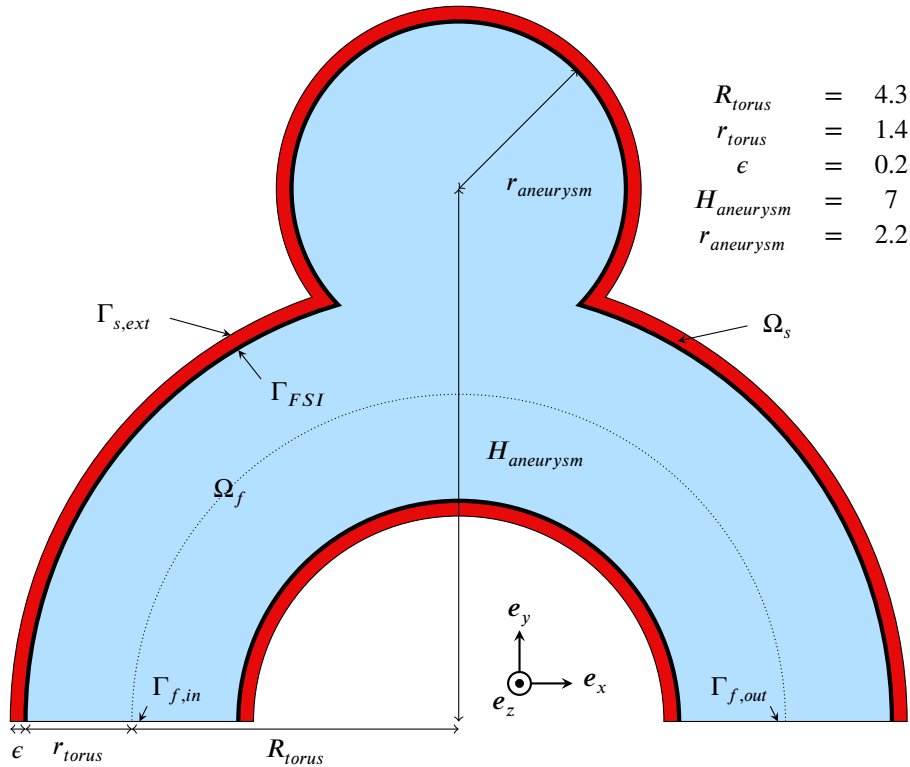

\begin{figure}
    \centering
    \begin{tabular}{c c c c}
        &\textit{S1} & \textit{S2} & \textit{S3}\\
        \toprule
        \rotatebox{90}{\hspace{15pt} \textit{IntrA dataset geometry}}
        &\includegraphics[width=0.3\linewidth]{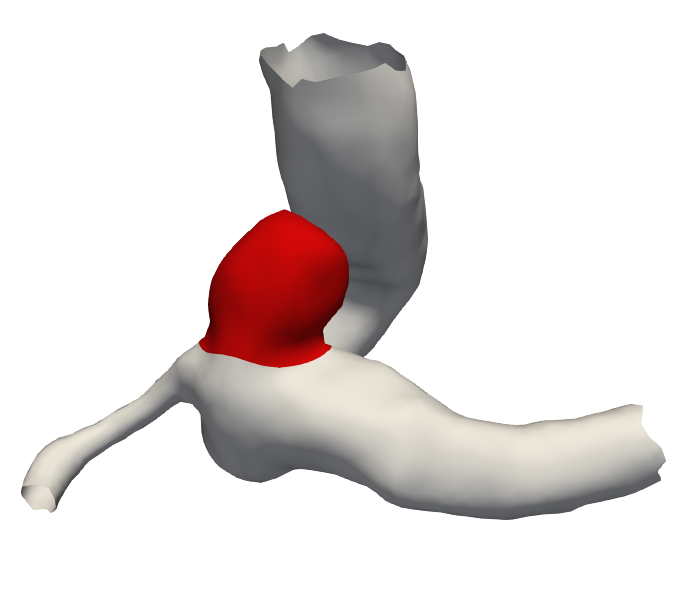}
        &\includegraphics[width=0.3\linewidth]{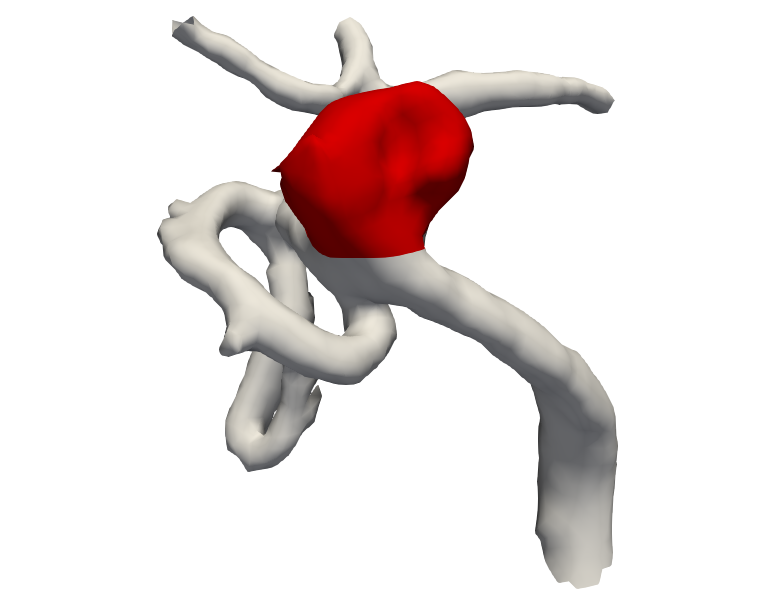}
        &\includegraphics[width=0.3\linewidth]{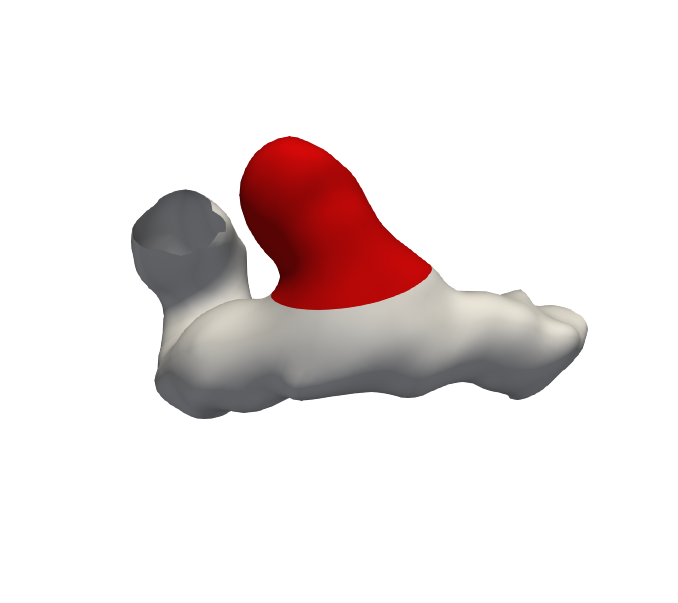}\\
        \noalign{\vskip 2mm}  
        \toprule
        \rotatebox{90}{\hspace{28pt} \textit{Adapted geometry}}
        &\includegraphics[width=0.3\linewidth]{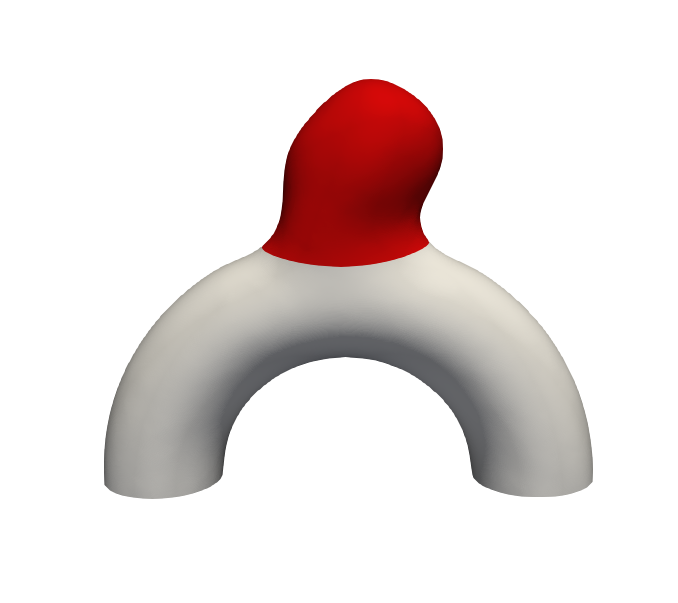}
        &\includegraphics[width=0.3\linewidth]{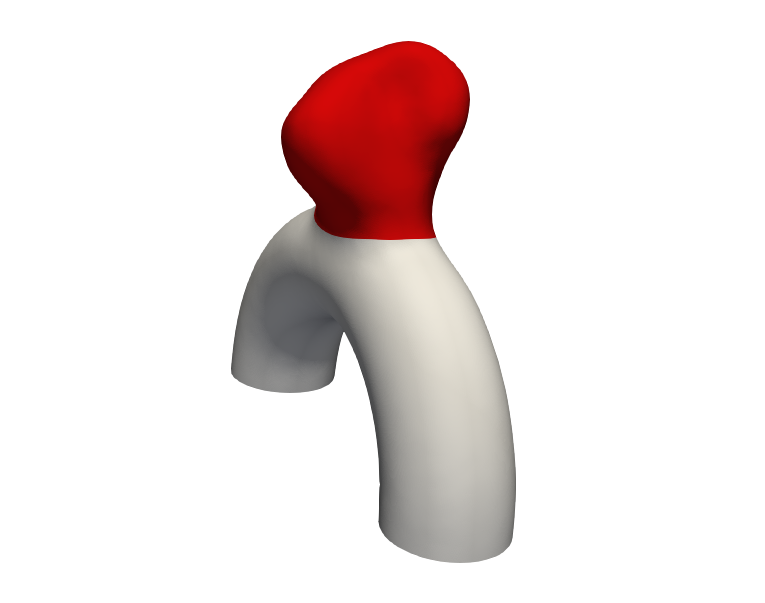}
        &\includegraphics[width=0.3\linewidth]{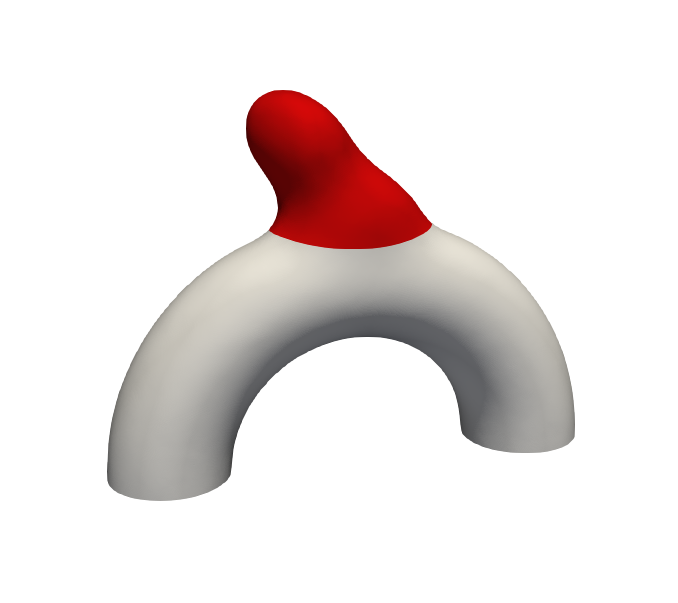}\\
        \toprule
    \end{tabular}
    \caption{Overview of the Specific cases adapted from the open-source \textit{IntrA} dataset \cite{INTRA}.}
    \label{fig:S_cases}
\end{figure}

A cut view of the proposed benchmark is shown in Figure \ref{fig:AnXplore_reference_definition}.
With an inlet diameter of $1.4\,mm$, the case mimics a simplified sidewall aneurysm located at the last segments (C6-C7) of a human Internal Carotid Artery (ICA) \cite{Carotid}.
These segments, named ophthalmic (C6) and communicating (C7), are fully located in the subarachnoidal space and constitute commonly known locations for sidewall IAs \cite{Day}.
The basic aneurysm bulge is designed as a perfect sphere ($4.4\,mm$ diameter) intersecting the toroidal geometry.
Proportions are inspired by previous work \cite{Ram+13, Has+05}.
The wall thickness is set to $200\,\mu m$ as will be discussed in the following section.
Given that it is both unrealistic and a potential source of problems in finite-element simulations, the singular sharp angles at the neck are smoothed, resulting in the 3D geometry visible in Figure \ref{fig:BC}.
In the following, we refer to this idealized geometry as R (for Reference).
Going a step further, three realistic bulge shapes taken from \cite{INTRA} have been employed, by adapting them manually to fit the neck of our idealized geometry.
The resulting shapes, presented in Figure \ref{fig:S_cases} will be referred to as S1, S2, S3 (for Specific).
The benchmark's modularity provides a functional exploration framework while preserving the biological and computational environment of IA modelling.

\subsubsection{Choice of physical parameters}
\label{sec:choice_physical_params}

\begin{figure}[h!]
    \centering
    
    \begin{tikzpicture}[scale=1]
    \node[inner sep=0pt] at (1,0)
    {\includegraphics[width=.3\textwidth]{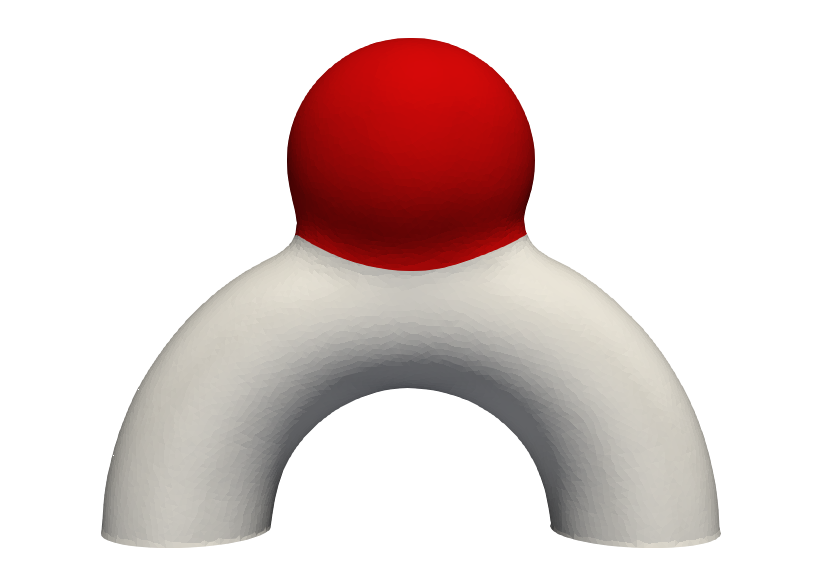}};

    \node at (-5.1,4) {
    \parbox{.3\linewidth}{
    \centering
    \begin{tabularx}{\linewidth}{
    	p{\dimexpr.4\linewidth-2\tabcolsep-1.3333\arrayrulewidth}
    	p{\dimexpr.5\linewidth-2\tabcolsep-1.3333\arrayrulewidth}
    	}
        \Xhline{4\arrayrulewidth}
    	$\rho_f$ & $10^3\,kg/m^3$\\
        \hline
        $\mu_0$ & $0.056\,Pa.s$\\
        \hline
        $\mu_\infty$ & $0.00345\,Pa.s$\\
        \hline
        $\lambda$ & $1.902\,s$\\
        \hline
        $n$ & $0.22$\\
        \hline
        $a$ & $1.25$\\
        \Xhline{4\arrayrulewidth}
    
    \end{tabularx}
    }
    };
    \node at (2,4) {
    \parbox{.45\linewidth}{
    \centering
    \begin{tabularx}{\linewidth}{
    	p{\dimexpr.25\linewidth-2\tabcolsep-1.3333\arrayrulewidth}
    	p{\dimexpr.75\linewidth-2\tabcolsep-1.3333\arrayrulewidth}
    	}
        \Xhline{4\arrayrulewidth}
        $\rho_{s,0}$ & $1.2\times10^3\,kg/m^3$\\
        \hline
    	$E_{aneurysm}$  & $1.38$ (R1) / $0.69$ (R2) / $0.35$ (R3) MPa\\
        \hline
        $E_{artery}$ & $6.9\,$MPa\\
        \hline
        $\nu$ & $0.45$\\    
        \Xhline{4\arrayrulewidth}
        $P_0$ & $-3.683\,kPa$\\
        \hline
        $R_d$ & $2.246\,kPa.s.mL^{-1}$\\
        \Xhline{4\arrayrulewidth}
    
    \end{tabularx}
    }
    };
    \draw [](2.5,-1.7) -- (2.5,-2.1);
    \draw [](3.6,-2.1) -- (2.5,-2.1);
    \draw (3.6,-2.2) rectangle (4.1,-2);
    \node at (3.85,-1.7) {$R_d$};
    \draw [](4.1,-2.1) -- (4.4,-2.1);
    \draw [](4.4,-2) -- (4.4,-2.2);
    \draw [](4.4,-2) -- (4.5,-2.05);
    \draw [](4.4,-2.1) -- (4.5,-2.15);
    \draw [](4.4,-2.2) -- (4.5,-2.25);

    \node[inner sep=0pt] at (-4.8,0)
    {\includegraphics[width=0.34\textwidth]{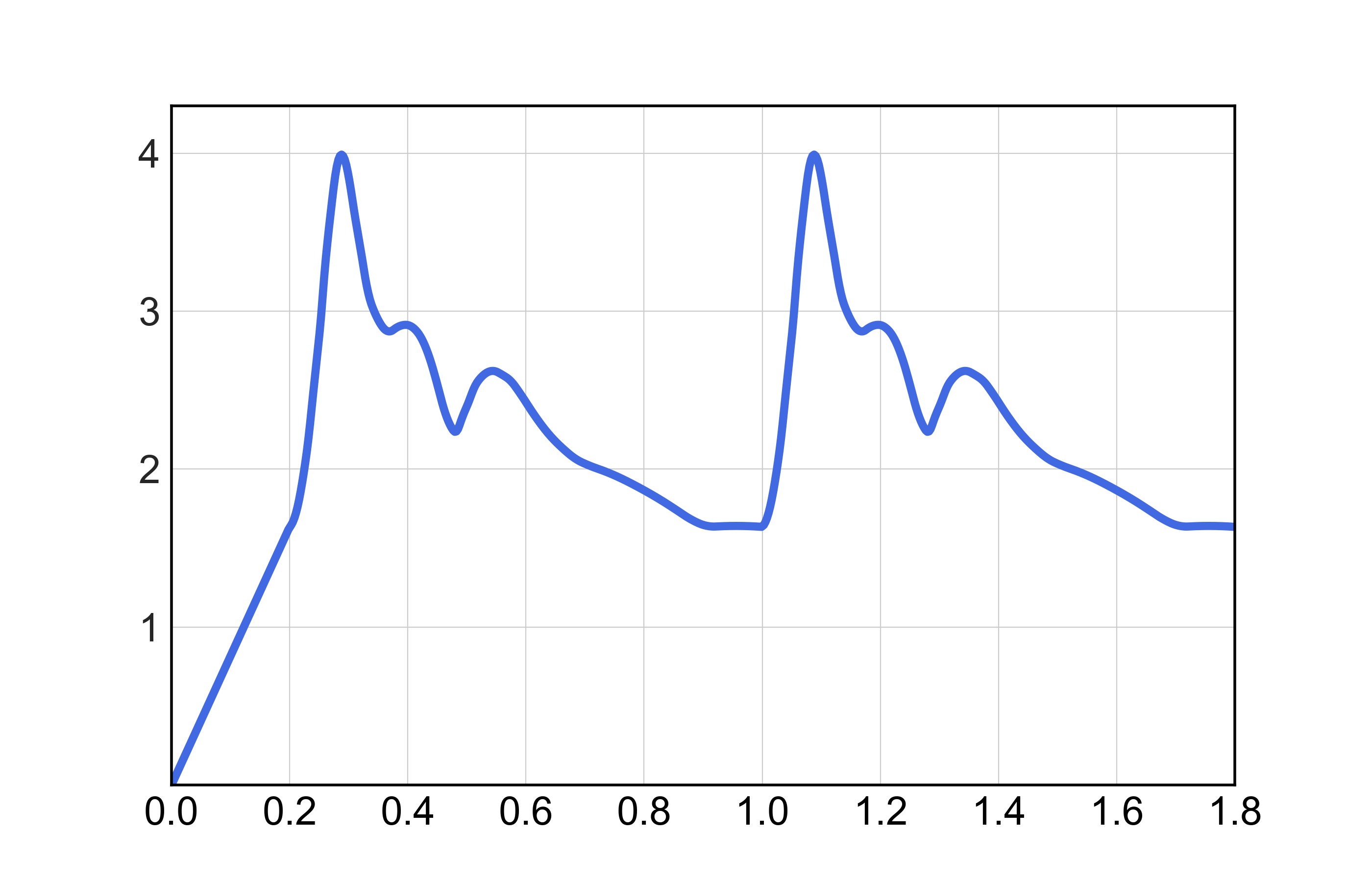}};

    \draw[-latex] (-2.2,-1.9) arc
    [
        start angle=-120,
        end angle=-52,
        x radius=1.5cm,
        y radius =1.5cm
    ] ;
    \draw [decorate, decoration = {brace}, ultra thick] (-8.01,2.55) --  (-8.01,5.4);
    \draw [decorate, decoration = {brace,mirror}, ultra thick] (6.2,3.55) --  (6.2,5.4);
    \draw [decorate, decoration = {brace}, ultra thick] (6.2,3.5) --  (6.2,2.55);
    \node[rotate=90] at (-8.5,4) {FLUID};
    \node[rotate=-90] at (6.68,4.5) {SOLID};
    \node[rotate=-90] at (6.68,3) {B.C.};

    \node at (-3.4,-1.15) {$\Delta t = 5\,ms$};
    \node at (-4.8,-2) {Time ($s$)};
    \node[rotate=90] at (-7.55,0) {Inflow rate ($mL.s^{-1}$)};
    \end{tikzpicture}
    \caption{Overview of the case settings and parameters. The generalized inflow waveform has been adapted from \cite{Ford2005Inflow}. The area coloured in red is where a lower solid stiffness is prescribed ($E_{aneurysm}$). B.C. stands for Boundary Conditions.}
    \label{fig:BC} 
\end{figure}
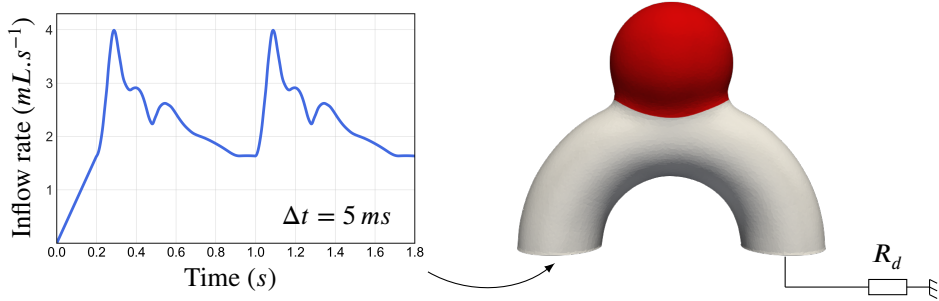

FSI in the context of IAs is undermined by uncertainty regarding pathological tissue characteristics.
Research effort has been carried out to measure the material properties of resected IA domes after clipping surgical operation \cite{Ceb+15, Rob+15, Lau+20}, analysing the alteration of collagen fiber architectures, and reporting a substantial scatter in ultimate stress and Young moduli between specimens.
Even on the surface of a single bulge, it has been shown that distinct regions can exhibit very different material properties \cite{Lau+20}.
IA wall thickness follows a similar trend.
Already in 1978, the authors of \cite{SO78} studied 45 clipped IAs and presented a diverse spectrum of wall thickness profiles.
This pathological tissue heterogeneity has been later quantified in several publications \cite{Lau+20, Aco+21}.
However, due to the lack of \textit{in-vivo} patient-specific data, most previous FSI studies modelling IAs employed constant wall thicknesses and isotropic mechanical properties \cite{Tor+08, Baz+10}, with the Neo-Hookean or Mooney-Rivlin models being commonly applied \cite{Tor+08, Tor+09, Baz+10, VoS+16}.
Even though implementing more realistic anisotropic models such as HGO \cite{HGO2000} is technically not a problem, applying it in a meaningful way is arduous, as literature does not provide insights into the preferred orientation of fibers in pathological tissue.
Furthermore, it has often been reported that, while hyperelastic modelling is required considering the observed deformation level of arterial walls, the kind of employed non-linearity has only little effect on the obtained results \cite{Baz+10, Tor+08}.
This comforts us in the choice of a Neo-Hookean model, as long as no more information is known about patient-specific tissue characteristics.
In addition, relying on a widely applied model like this one makes the case easier to reproduce.

Regarding the selected values of wall characteristics, a typical Young modulus of $E = 1\,$MPa (with Poisson ratio $\nu = 0.45$) is frequently reported  \cite{Tor+08, Tor+09, Baz+10, VoS+16} and prescribed wall thicknesses range from $200\,\mu m$ \cite{Ceb+14, Baz+10} to $350\,\mu m$ \cite{Val+13}, with some studies reducing it drastically at the aneurysm bulge \cite{Tor+09, Ceb+14}.
In our case, to circumvent the difficult choice of absolute parameter values based on experimental measurement and characterization, we decide to rely on the rare visual evidence of brain arterial wall movement.
As mentioned in the introduction, recent studies have reported IA pulsation observation through direct \textit{in-vivo} imaging \cite{Van+15, Hay+13, Zho+22, Sta+21}. 
We use these volume variation measurements as an approximate reference and adjust the wall stiffness to calibrate the amplitude of the movement.
As no guideline exists to vary the thickness and stiffness of the walls in a meaningful way, we prefer solely altering the stiffness, viewed as a global wall structural resistance.
In the following, we explore 3 configurations (R1, R2, R3) with $E_1 = 1.38\,$MPa (Lame coefficient $\mu_1 = 0.4\,$MPa), $E_2 = E_1/2$, $E_3 = E_2/2$.
These configurations should result in pressure-induced volume changes that approximately cover the spectrum of reported values \cite{Sta+21}.
To draw the focus solely on the aneurysm bulge interaction with the flow, we decide to drastically raise the stiffness of the artery to $E_{artery} = 5E_1$.
The bulge area, where the lower stiffness is prescribed, is shown in red in Figure \ref{fig:BC}.
Note that the intermediary stiffness $E_2$ is employed for the Specific shapes (S1-3).

Thanks to abundant literature, choosing fluid properties stands as less problematic.
A shear-thinning Carreau-Yasuda rheology model is implemented as in Eq. \eqref{eq:carreau}.
The employed parameters based on \cite{Robertson2009} are summarized in Figure \ref{fig:BC}.
\begin{equation}
    \mu(\dot{\gamma}) = \mu_\infty + (\mu_0 - \mu_\infty)\left( 1+(\lambda\dot{\gamma})^a \right)^{(n-1)/a}
    \label{eq:carreau}
\end{equation}
\\
\\

\subsubsection{Boundary conditions}

In the fluid, inlet velocities and outlet pressure conditions are prescribed as follows:

\begin{align}
    \begin{split} \label{equation:BC}
        \vv(\bm{x},t) &= V(t) \left( 1-  \left(\frac{||\bm{x}-R_{torus}||}{r_{torus}} \right)^2 \right)\,\bm{e}_y, \quad \bm{x} \in \Gamma_{f,in}, \\
        P(t) &= P_0 + R_d \int_{\Gamma_{f,out}} \vv(\bm{x},t) \cdot (-\bm{e}_y)\,d\Gamma,  \quad \text{on} \: \Gamma_{f,out}.
    \end{split}
\end{align}

Where $V(t)$ has been built based on the waveform plotted in Figure \ref{fig:BC}, corresponding to an averaged internal carotid pulse reported by \cite{Ford2005Inflow}.
It has been scaled with respect to the inlet diameter to reach realistic flow rates and prefixed with a $0.2\,s$ linear ramp for a smoother initialization.

At the outflow, contrary to regular CFD simulations, the absolute value of the pressure is of major importance.
To reach plausible deformations of the aneurysmal membrane, physiological pressures have to be applied.
Pressures in the vascular system dwell between $80$ and $120\,mmHg$ for healthy patients.
These pressure variations occur over a cardiac cycle in the system and result from the hydraulic resistance of the posterior vasculature ($R_d$), mostly imputable to brain capillaries.
To account for that, the outflow pressure is scaled with respect to the flow rate (i.e. adjusting $R_{d}$) to keep the pressure in the system between given bounds, similar to \cite{Tor+08}.
Considering the lack of information regarding the non-linear behaviour of pathological aneurysmal tissue, and for the sake of easier reproducibility, we decide to vary the outflow pressure between $0$ and $40\,mmHg$ rather than pre-stressing the diastolic structure as it has been done by \cite{Baz+10}.
Thus, we set $P_0 = -3.683\,kPa$ and $R_d = 2.246\,kPa.s.mL^{-1}$.

For the solid, boundary conditions are straightforward.
Nodes situated on the inflow/outflow plane ($y=0$) are kept fixed, whereas a traction-free condition is prescribed on $\Gamma_{s,ext}$.

\subsubsection{Quantities of interest}
\label{sec:qoi}

One of the major goals of this practical benchmark lies in the ability to assess the sensitivity of haemodynamics to FSI modelling.
The most widely used metric for rupture risk assessment of IAs is the Wall Shear Stress (WSS) applied by the blood flow in the bulge. 
WSS is associated with remodelling pathways of IAs \cite{Men+14, Fro+04, Men+07, Ceb+14} leading to their formation and growth, due to the interaction between abnormal blood flow and the endothelial cells of the vessels \cite{MAl99}.
As multiple definitions exist, the one employed in this work is provided in Equation \ref{eq:WSS_def2} for the sake of reproducibility:

\begin{gather}
\bm{\tau}_{WSS} = \bm{n}\times \left[\left(\sigmaF\cdot\bm{n}\right)\times\bm{n}\right] = \sigmaF\cdot\bm{n} - \left[\left(\sigmaF\cdot\bm{n}\right)\cdot\bm{n}\right] \bm{n}
    \label{eq:WSS_def2}
\end{gather}

Where $\bm{n}$ is the unit normal vector at the wall and $\sigmaF$ the Cauchy stress tensor defined as: $\sigmaF = - p_f \mathbf{I} + \mu_f (\nabla \vv + \nabla^T \vv)$.
This vectorial definition allows computing another important metric known as the Oscillatory Shear Index (OSI):

\begin{equation}
    \text{OSI} = \frac{1}{2}\left( 1 - \frac{ ||\int_{t_0}^{t_0+T} \bm{\tau}_{WSS}~dt|| }{ \int_{t_0}^{t_0+T} ||\bm{\tau}_{WSS}||~dt}\right)
    \label{eq:osi}
\end{equation}

In the presented results, we will emphasize these two indicators and record them over the second cardiac cycle only (from $t_0 = 1\,s$ to $t_0+T = 1.8\,s$) in order to limit any transient effect associated with the initial flow development. 
WSS will always be reported as a scalar quantity which corresponds to the Euclidian norm of $\bm{\tau}_{WSS}$.
WSS will be recorded at systole ($t = 1.08\,s$) and averaged in time to yield the Time-Averaged WSS (TAWSS).

\subsubsection{Meshing}
\label{sec:meshing}
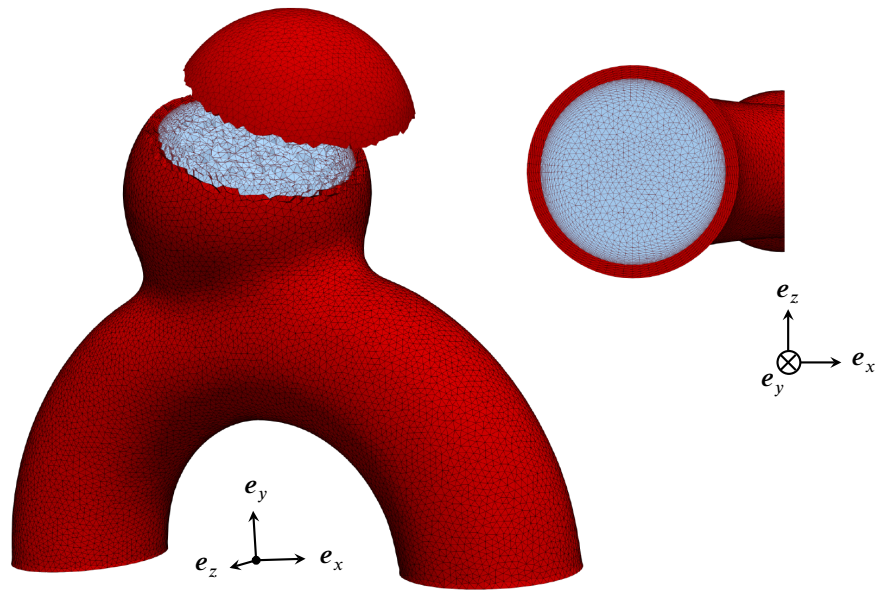
\begin{figure}[ht!]
    \centering
    
    \begin{tikzpicture}[scale=1]

    \node[inner sep=0pt] at (-0.2,-6.3)
    {\includegraphics[width=.65\textwidth]{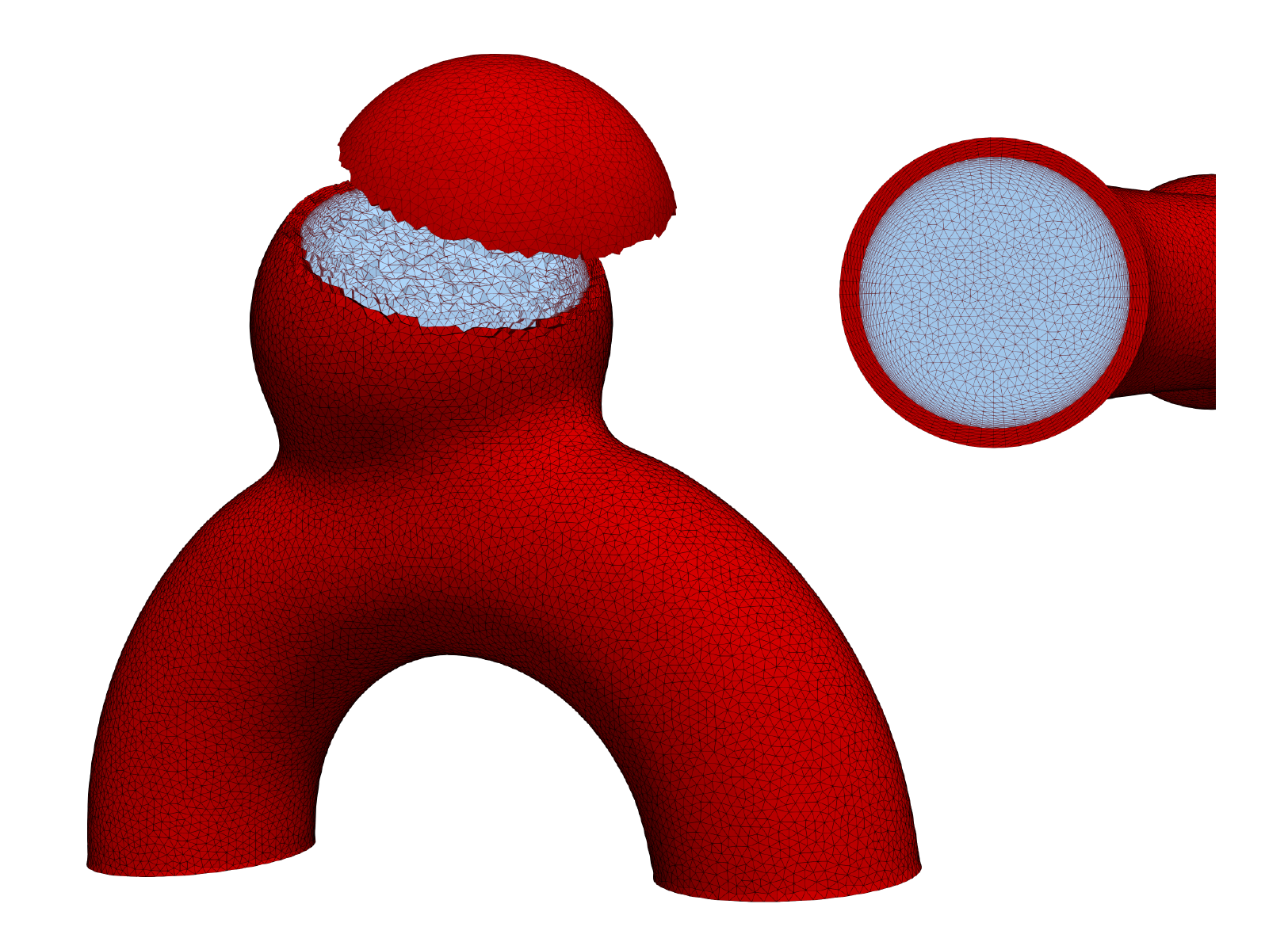}};
    \draw [-stealth, thick](-2,-9.8) -- (-1.35,-9.78);
    \draw [-stealth, thick](-2,-9.8) -- (-2.04,-9.15);
    \draw [-stealth, thick](-2,-9.8) -- (-2.35,-9.9);
    \draw[fill=black, ultra thick] (-2,-9.8) circle (0.03cm);
    \node at (-2,-8.9) {$\bm{e}_y$};
    \node at (-1,-9.8) {$\bm{e}_x$};
    \node at (-2.65,-9.9) {$\bm{e}_z$};

    \draw [-stealth, thick](5,-7.2) -- (5.7,-7.2);
    \draw [-stealth, thick](5,-7.2) -- (5,-6.5);
    \draw [fill=white, thick](5.01,-7.2) circle (0.15cm);
    \draw [thick](4.9,-7.1) -- (5.1,-7.3);
    \draw [thick](4.9,-7.3) -- (5.1,-7.1);
    \node at (5,-6.3) {$\bm{e}_z$};
    \node at (6,-7.2) {$\bm{e}_x$};
    \node at (4.8,-7.5) {$\bm{e}_y$};
    
    \end{tikzpicture}
    \caption{Views of the employed meshes with a focus at the inlet.}
    \label{fig:MeshView}
\end{figure}

The primary variables computed when solving the Navier-Stokes equation are velocity and pressure.
In our case, the velocity is obtained as a P1 field (see next Section) and gradients are computed through post-processing.
As a consequence, the mesh resolution is of major importance especially in the vicinity of the walls, to resolve high gradients properly.
Therefore, we use a boundary layer in the fluid domain as shown in Figure \ref{fig:MeshView}.
We employ a geometrical progression (factor $\alpha = 1.2$) between successive layers, a minimal element thickness of $10\mu m$ and a total boundary layer size of $0.3\,mm$.
The isotropic element size of the core mesh is set to $0.12\,mm$, and the solid thickness is divided into 6 equal layers of $0.033\,mm$.
This results in a mesh composed of $1.2\,M$ and $0.5\,M$ elements, for the fluid and the solid of the reference case (R), respectively.
All the meshes have been generated with the Gmsh \cite{gmsh} python package and are available on \href{https://github.com/aurelegoetz/AnXplore}{GitHub}.

\subsection{Modelling the physics}
\subsubsection{Fluid solver}

When simulating haemodynamics in compliant arteries, keeping a fitted fluid-structure coupling interface is decisive in obtaining precise WSS estimates (see Section \ref{sec:qoi}).
Furthermore, as mesh deformations remain moderate, an Arbitrary Lagrangian-Eulerian (ALE) \cite{ALEorigin} description appears as the most practical approach to employ.
Thus, we define $\Omega_{f,t} \subset \: {\rm I\!R}^n$ the fluid spatial domain at time $t\:\in\:[0,T]$, with $n$ the spatial dimension, and $\boldsymbol{\psi}$ the ALE mapping from $\Omega_{f,0}$ to $\Omega_{f,t}$.
The associated relative velocity is denoted $\vm$.
Let $\Gamma_f$ be the boundary of $\Omega_f$.
We consider the mixed formulation in velocity $\vv$ and pressure $p_f$ of the transient incompressible Navier-Stokes equations given by:
\begin{equation}
    \rho_f \partial_t \vv + \rho_f ((\vv - \vm) \cdot \nabla) \:\vv - \nabla \cdot \sigmaF = \bm{f}, \quad \text{in} \: \Omega_{f,t}.
    \label{equation:NS}
\end{equation}
\begin{equation}
    \nabla \cdot \vv = 0, \quad \text{in} \: \Omega_{f,t}.
    \label{equation:continuity}
\end{equation}

where $\rho_f$ is the fluid mass density and $\bm{f}$ the source term.
We rely on a P1-P1 finite element discretization for solving the mixed-formulation (Eqs. \eqref{equation:NS}, \eqref{equation:continuity}), combined with a Variational Multiscale-type (VMS) method as described in \cite{VMS}. 
This method ensures accuracy and stability \cite{babuvska1971error} even for convection-dominated flow by enriching both velocity and pressure with residual-based subscales.

In the ALE framework, the convective velocity is altered by the mesh velocity $\vm$, which tracks the movement of coupling interfaces ($\Gamma_{FSI}$).
The adaptive mesh displacement allows keeping boundaries fit and consequently saves the cost of interpolating between subdomains.
A $C_2$-smooth $\vm$ field can be obtained, for instance, by solving the diffusion equation \cite{HABCHI2013306}:

\begin{align}
\begin{split} \label{equation:diffusion}
    \nabla \cdot (\gamma\nabla \vm) &= 0, \quad \text{on} \: \Omega_{f},\\
    \vm &= \partial_t \uu,  \quad \text{on} \: \Gamma_{FSI},\\
    \vm &= 0,  \quad \text{on} \: \Gamma_f \setminus \Gamma_{FSI},
    \end{split}
\end{align}
where $\uu$ stands for the displacement of the solid interacting with the considered fluid domain.
The diffusion coefficient $\gamma$ is taken to be the squared inverse distance to the interface $\Gamma_{FSI}$, in order to better share the mesh deformation on the entire grid and to keep the boundary layer mesh as intact as possible.
Many other solutions exist and the interested reader can refer to \cite{Sha20, ElasticMeshMove}.

\subsubsection{Solid solver}
\label{sec:solid}

The compliant arterial tissue is modelled using the Lagrangian equations of solid dynamics.
Let $\Omega_{s,0}$ and $\Omega_{s,t}$ define the initial and current solid spatial domains, with $\bm{\phi}$ the mapping between the two domains.
We should distinguish here between the material Lagrangian coordinate $\mathbf{X}$, and the updated Lagrangian coordinate $\mathbf{x}$.
The displacement of a solid particle is given by $\mathbf{u}= \mathbf{x} - \mathbf{X} = \boldsymbol{\phi}(\mathbf{X},t) - \mathbf{X}$ and the deformation gradient defined as $\mathbf{F} =  \nabla_{\mathbf{X}} \boldsymbol{\phi}$.
The Jacobian determinant is thus $J = \text{det}[\mathbf{F}]$.
The momentum and continuity equations for solid dynamics are specified as:
\begin{equation}
\rho_s \partial_{tt} \uu - \nabla_{\mathbf{x}} \cdot \: \sigmaS = 0, \quad \text{in} \: \Omega_{s,t}.
\end{equation}
\begin{equation}
\rho_s J = \rho_{s_0}, \quad \text{in} \: \Omega_{s,t}.
\end{equation}

Where $\rho_s$, $\ddot{\uu}$, and $\sigmaS$, designate the solid density, the second material derivative of the displacement and the symmetric Cauchy stress tensor, respectively.

For modelling the intrinsically hyperelastic nature of arterial tissue \cite{Baz+10,Tor+08,HGO2000}, we rely on the Helmholtz free energy formalism.
Let $\bm{C}$ denote the right Cauchy-Green strain tensor given by $\bm{C} = \bm{F}^T\bm{F}$ and $\bm{S} =  J \bm{F}^{-1} \sigmaS \bm{F}^{-T}$ the second Piola--Kirchhoff stress tensor.
The Helmholtz free energy function $\boldsymbol{\Psi} (\textbf{C})$ is defined by:

\begin{equation}
\textbf{S}=2\partial_{\textbf{C}} \Psi (\textbf{C}).
\end{equation}

This free energy function is decomposed into its volumetric and deviatoric contributions, leading to the classical split:

\begin{equation}
\Psi ( \textbf{C} ) = U(J) + W( \bar{\textbf{C}} ).
\end{equation}

Where  $J = \sqrt{\text{det}[\textbf{C}]}$, and $\bar{\textbf{C}} = J^{-\frac{2}{3}} \textbf{C}$ is the volumetric/deviatoric part of $\textbf{C}$.

We Consider a Neo-Hookean and a Simo--Taylor \cite{SimoTaylor} volumetric model, which yields:

\begin{equation}
U( J ) = \frac{1}{4} \kappa ( J^2 - 1 ) - \frac{1}{2} \kappa \text{ln} J,
\end{equation} 

\begin{equation}
W ( \bar{\textbf{C}} ) = \frac{1}{2} \mu_s ( \text{tr}[\bar{\textbf{C}}] - 3 )  =\frac{1}{2} \mu_s ( \bar{\mathbf{I}_1} - 3 ).
\end{equation}

where $\kappa$ and $\mu_s$ are material properties,and ${ \mathbf{I}_1}=\text{tr}[\bar{\textbf{C}}]$ is the first Cauchy-Green invariant.
The Cauchy stress tensor can similarly be split into its deviatoric and volumetric parts, which gives:

\begin{equation}
\sigmaS = p_s \mathbf{I} + \text{dev}[\sigmaS] .
\end{equation}

\begin{equation}
p_s = 2 J^{-1} \textbf{F} \partial_\textbf{C} U(J) \textbf{F}^T = U'(J) = \frac{1}{2} \kappa (J + J^{-1}),
\end{equation}

\begin{equation}
\text{dev}[\sigmaS] = 2 J^{-1} \textbf{F} \partial_\textbf{C} W(\bar{\textbf{C}}) \textbf{F}^T = \mu_s J^{-\frac{5}{3}} \text{dev}[ \textbf{FF}^T].
\end{equation}

The final system of equations to be solved is given by:

\begin{equation}
\rho_s \partial_{tt} \uu - \nabla_{\mathbf{x}} p_s - \nabla_{\mathbf{x}} \cdot \: \text{dev}[\sigmaS] = 0, \quad \text{in} \: \Omega_s.
\end{equation}

\begin{equation}
  \nabla_{\mathbf{x}} \cdot \uu - \frac{1}{\kappa} p_s  = 0, \quad \text{in} \: \Omega_s.
\end{equation}

As for the fluid, we rely on a stabilized method for tackling the solid finite-element problem.
More details about the method and implementation are given in \cite{NEMER2021113923}. 

\subsubsection{Coupling}

Dynamic and kinematic coupling conditions must be enforced at the fluid-solid interface, whose normal field is denoted $\bm{n}$:

\begin{align}
    \begin{split} \label{equation:coupling}
        \vv &= \partial_t \uu, \quad \text{on} \: \Gamma_{FSI}, \\
        \sigmaF \cdot \bm{n} &= \sigmaS \cdot \bm{n}, \quad \text{on} \: \Gamma_{FSI}.
    \end{split}
\end{align}

The interface continuity constraints are imposed using a partitioned, iterative scheme \cite{FELIPPA198061} with the convergence criterion \eqref{equation:coupling}.
This sub-iterative process appears especially crucial when fluid and solid densities come close or when dealing with slender solid geometries, in order to not suffer from coupling instabilities commonly known as \textit{added-mass effect} \cite{CAUSIN20054506, Forster}.
We employ a classic Dirichlet-to-Neumann coupling, where the velocity of the solid is imposed as a Dirichlet condition at the fluid boundary, while the normal fluid stress is enforced as a Neumann condition on the solid.

Reaching FSI convergence can be mathematically viewed as finding a fixed point of the solid and fluid operators' composition ($\mathcal{S} \circ \mathcal{F}$).
We define the FSI residual as follows:
\begin{equation}
    \textbf{r}_{t}^k = \uu_{t}^k - \tilde{\uu}_{t}^k = \mathcal{S} \circ \mathcal{F} (\tilde{\uu}_{t}^k) - \tilde{\uu}_{t}^k,
\end{equation}
where $\tilde{\uu}_{t}^k$ is the predicted displacement of the solid used at sub-increment $k$.
Time will only be incremented after this fixed point has been reached with a given tolerance: $||\textbf{r}_{t}^k|| < tol_{FSI}$ ($||.||$ being the euclidian norm scaled with the number of nodes in the mesh).
For the proposed test case (R or S), this tolerance is set to $10^{-4}\,mm$.
For the fixed-point algorithm to converge quicker, under-relaxation is widely employed \cite{HABCHI2013306, BREUER2012107, Eken}.
It consists in using only a fraction of the algorithm's new solution ($\uu_{t}^k$) for building the next guess.
Mathematically, relaxing the solution with a relaxation parameter $\omega$ can be written as:
\begin{equation}
    \tilde{\uu}_{t}^{k+1} = \tilde{\uu}_{t}^{k} + \omega(\uu_{t}^k-\tilde{\uu}_{t}^{k}) = \tilde{\uu}_{t}^{k} + \omega \textbf{r}_{t}^k
    \label{equation:relaxation}
\end{equation}
From within the large family of relaxation methods \cite{Kuttler}, we choose the momentum accelerated Aitken $\Delta2$ scheme for its well-studied properties and extensive use in the community.
The dynamic relaxation parameter $\omega_{t}^k$ can be assessed at every subincrement $k$ by using the following formula \cite{Kuttler}:

\begin{equation}
    \omega_{t}^{k} = -\omega_{t}^{k-1} \frac{(\textbf{r}_{t}^{k-1})^T (\textbf{r}_{t}^{k} - \textbf{r}_{t}^{k-1})}{||(\textbf{r}_{t}^{k} - \textbf{r}_{t}^{k-1})||^2}
    \label{equation:Aitken}
\end{equation}

However, one should note that Eq. \eqref{equation:relaxation} cannot be used to initialize the predicted displacement $\tilde{\uu}_{t}^{1}$ when starting a new timestep, as no previous subincrement exists.
As a result, a linear predictor is employed, based on previous converged displacement.
Similarly, Eq. \eqref{equation:Aitken} can only be used as from the end of the second subincrement.
Before that, a fixed value $\omega_0 = 0.1$ is employed.

\section{Results}
\subsection{Validation with the pressure wave benchmark}

\begin{figure}[h!]
    \centering
    
    \begin{tikzpicture}[scale=2.2]
    \draw[fill=bluefluid] (0,-0.5) rectangle (5,0.5);
    \draw[fill=redsolid] (0,0.5) rectangle (5,0.63);
    \draw[fill=redsolid] (0,-0.63) rectangle (5,-0.5);
    \draw [ultra thick](0,0.5) -- (5,0.5);
    \draw [ultra thick](0,-0.5) -- (5,-0.5);
    \draw [ultra thick, densely dotted](0,0.5) -- (0,0.63);
    \draw [ultra thick, densely dotted](0,-0.5) -- (0,-0.63);
    \draw [ultra thick, densely dotted](5,0.5) -- (5,0.63);
    \draw [ultra thick, densely dotted](5,-0.5) -- (5,-0.63);
    \node at (2.5,0) {$\Omega_f$};
    \draw [-stealth](4.5,0.8) -- (3.5,0.56);
    \draw [-stealth](4.5,0.8) -- (3.5,-0.59);
    \node at (4.67,0.8) {$\Omega_s$};
    \draw[fill=black, ultra thick] (2.5,0.5) circle (0.015cm);
    \node at (2.58,0.38) {$A$};
    \draw [<->](-0.1,-0.5) -- (-0.1,0.5);
    \node at (-0.2,0) {$2r$};
    \draw [<->](-0.1,0.5) -- (-0.1,0.63);
    \node at (-0.2,0.55) {$\epsilon$};
    \draw [<->](-0.1,-0.5) -- (-0.1,-0.63);
    \node at (-0.2,-0.55) {$\epsilon$};
    \draw [<->](0,-0.7) -- (5,-0.7);
    \node at (2.5,-0.8) {$L$};
    \draw [-stealth](0.2,0.05) -- (0,0);
    \node at (0.45,0.05) {$\Gamma_{f,in}$};
    \draw [-stealth](4.8,0.05) -- (5,0);
    \node at (4.5,0.05) {$\Gamma_{f,out}$};
    \draw [-stealth](1.2,0) -- (0.7,0.5);
    \draw [-stealth](1.2,0) -- (0.7,-0.5);
    \node at (1.5,0) {$\Gamma_{FSI}$};
    \draw [-stealth](2,0.8) -- (1.7,0.63);
    \draw [-stealth](2,0.8) -- (1.7,-0.63);
    \node at (2.27,0.8) {$\Gamma_{s,ext}$};
    \draw [ultra thick, densely dotted](0.2,0.82) -- (0.33,0.82);
    \node at (0.69,0.8) {$\Gamma_{s,clamped}$};
    \node at (5.65,0){
    \begin{tabular}{cc}
        $r =$  & 5\\
        $L =$ & 50\\
        $\epsilon =$ & 1\\
    \end{tabular}
    };
    \node at (2.5,-1.75) {
    {\includegraphics[width=0.69\linewidth]{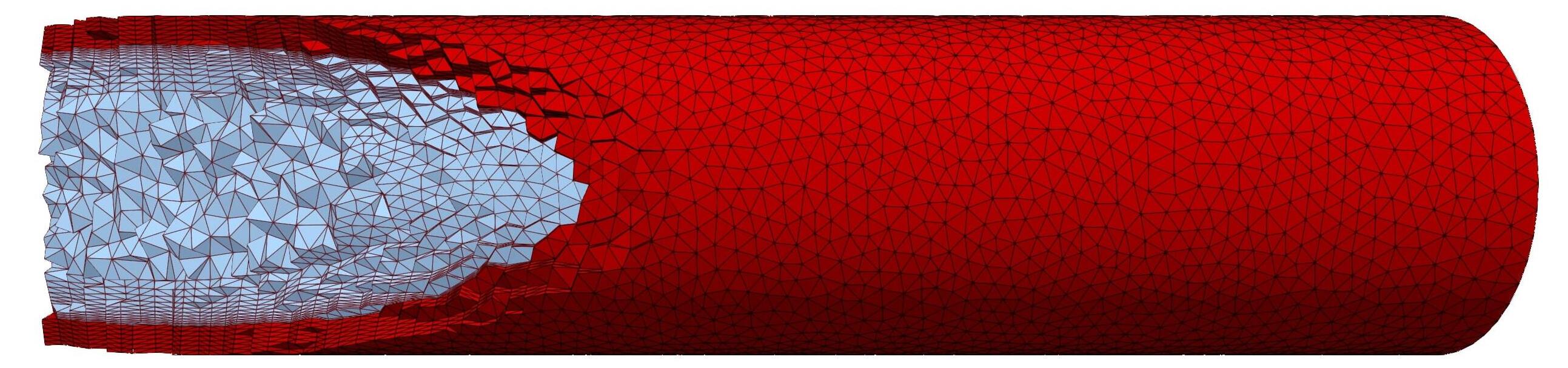}}
    };
    \end{tikzpicture}
    \caption{Geometry of the pressure wave benchmark \cite{formaggia} (dimensions in mm) along with a view of the M mesh (see Table \ref{tab:meshesPW3}).}
    \label{fig:PW3_definition}
\end{figure}
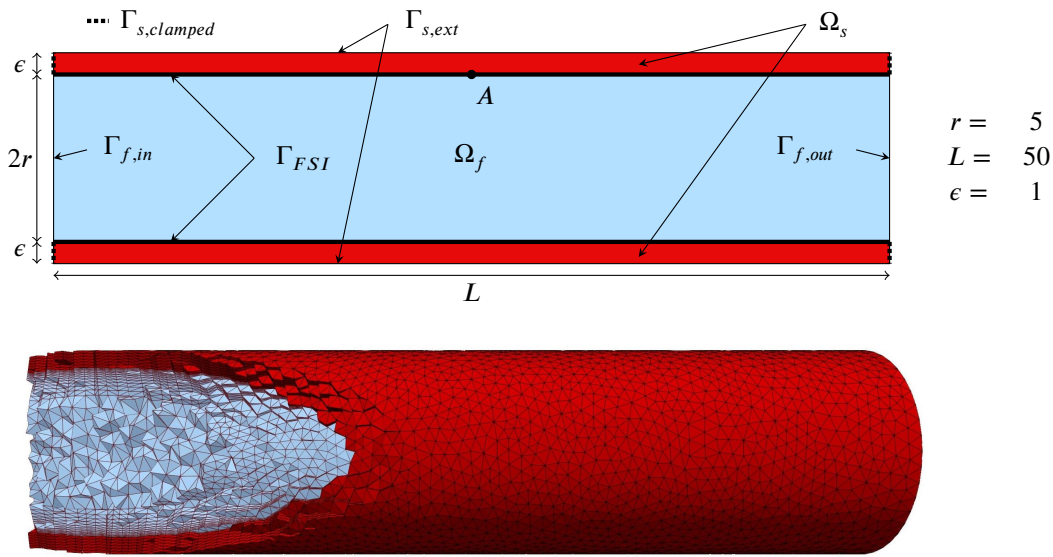

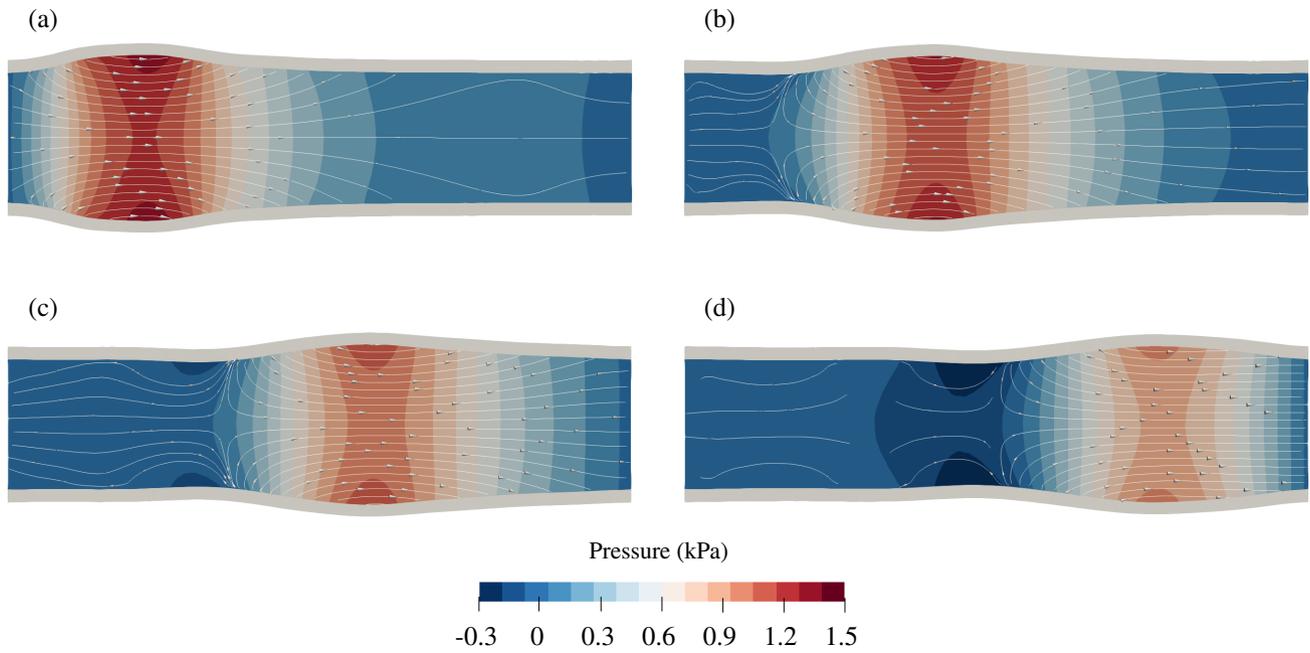
\begin{figure}[h!]
    \centering
    \begin{tikzpicture}
    \node at (0,0) {
        \includegraphics[width=1\linewidth]{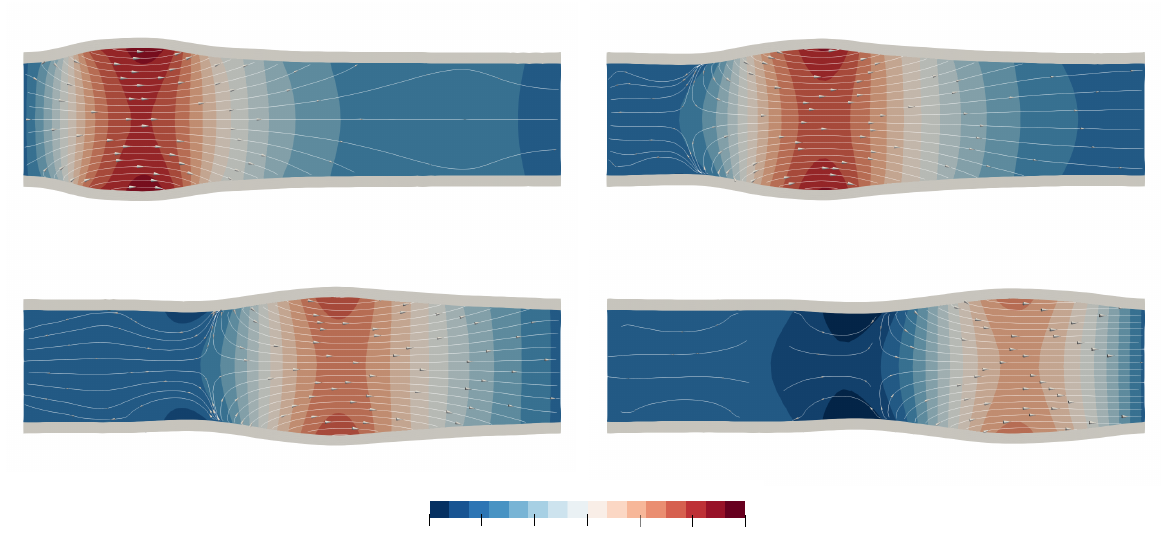}
    };

    \node at (-8.1,4) {(a)};
    \node at (-8.1,0.2) {(c)};
    \node at (0.8,4) {(b)};
    \node at (0.8,0.2) {(d)};
    
    \node[] at (0,-3) {\fontsize{9}{12}\selectfont Pressure (kPa)};
    \node at (-2.4,-4.1) {-0.3};
    \node at (2.4,-4.1) {1.5};
    \node at (1.6,-4.1) {1.2};
    \node at (-1.6,-4.1) {0};
    \node at (0.8,-4.1) {0.9};
    \node at (-0.8,-4.1) {0.3};
    \node at (0,-4.1) {0.6};

    \end{tikzpicture}
    \caption{Middle cross-section view of pressure contours along with streamlines for different times: (a) $t=4\,ms$, (b) $t=6\,ms$, (c) $t=8\,ms$ and (d) $t=10\,ms$. Displayed results correspond to the medium mesh (M in Table \ref{tab:meshesPW3}). Structural displacements have been amplified by a factor 10.}
    \label{fig:PW3_pressure_streamlines}
\end{figure}

\begin{figure}[h!]
    \centering
    \begin{tikzpicture}[scale=2.2]
    \node at (0,0) {
    {\includegraphics[width=1\linewidth]{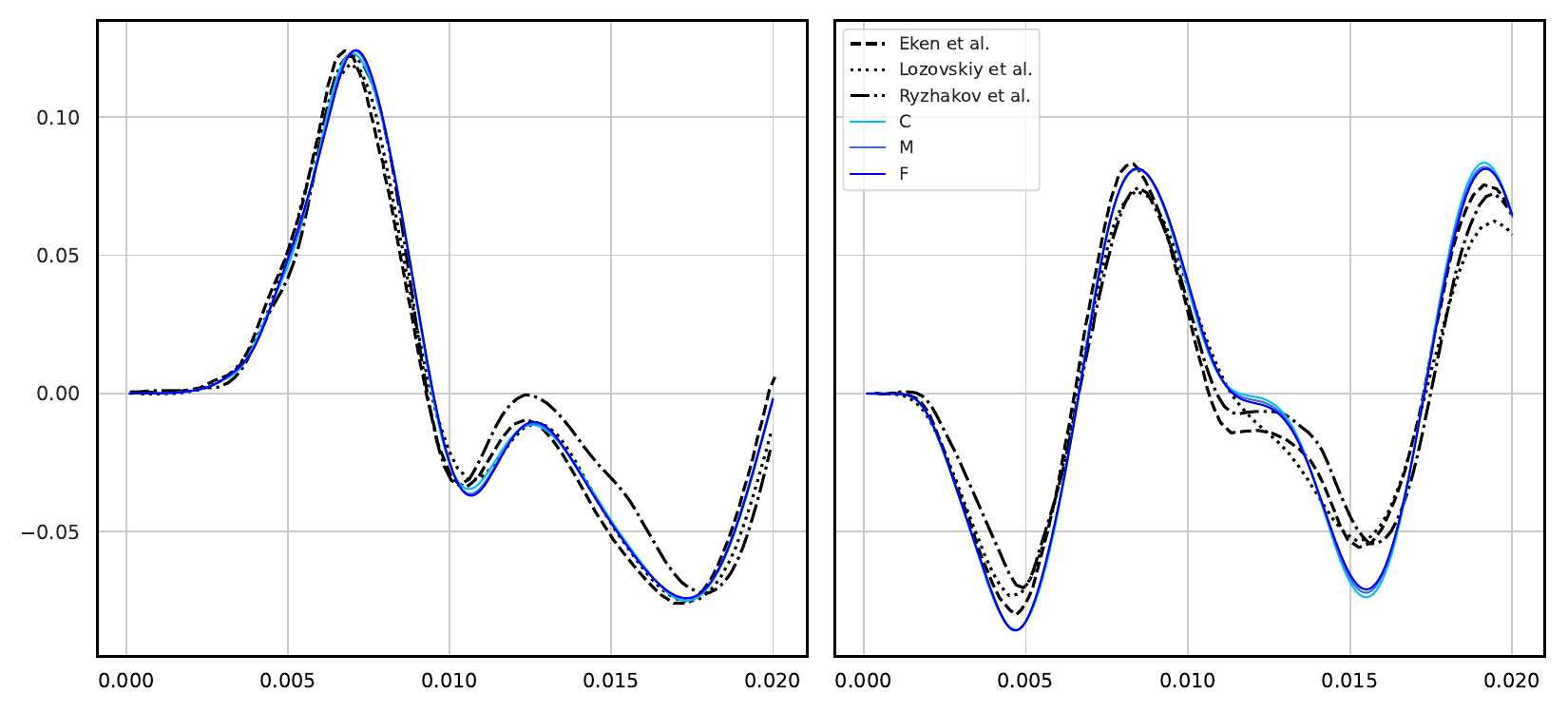}}
    };
    \node[rotate=90] at (-4,0) {Displacement (mm)};
    \node at (-1.65,-1.9) {Time (s)};
    \node at (2.15,-1.9) {Time (s)};
    \node at (-0.22,1.35) {\textbf{(a)}};
    \node at (3.6,1.35) {\textbf{(b)}};
    \end{tikzpicture}
    \caption{Radial (a) and longitudinal (b) displacements of point A (see Figure \ref{fig:PW3_definition}) for the three meshes introduced in Table \ref{tab:meshesPW3}.
    Results from \cite{Eken}, \cite{LOZOVSKIY} and \cite{Ryzha} have been reproduced for comparison.}
    \label{fig:PW3_results}
\end{figure}
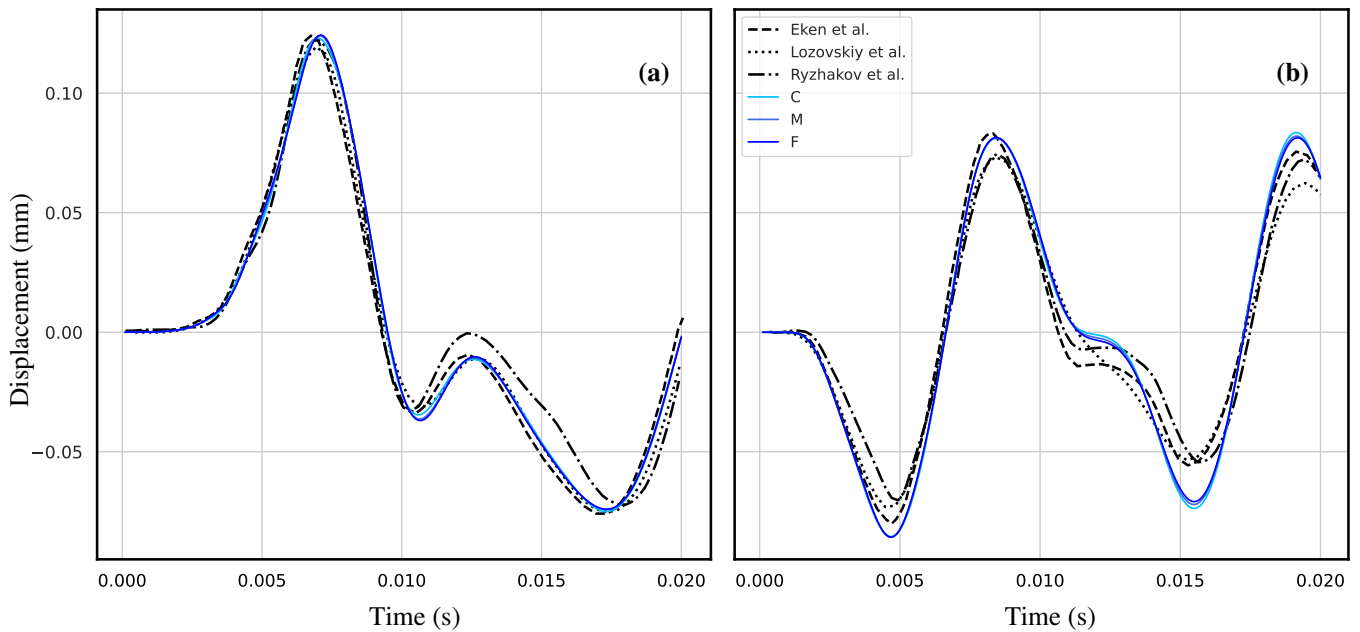

To assess the accuracy and convergence of the employed FSI method, the well-known pressure wave benchmark \cite{formaggia} is implemented.
This FSI case has been reproduced in several studies \cite{MALAN2013426, Ryzha, Eken, LOZOVSKIY} and remains the only widespread benchmark that models the coupling between arterial tissue and blood flow.
Geometrical features are described in Figure \ref{fig:PW3_definition}.
A fluid kinematic viscosity of $\nu_f\,$\,$=$\,$10^{-6}\,m^2/s$ is set, along with fluid and solid densities of $\rho_f\,$\,$=$\,$10^3\,kg/m^3$, $\rho_{s,0}\,$\,$=$\,$1.2\times10^3\,kg/m^3$, respectively.
In the original work of Formaggia et al. \cite{formaggia}, a Saint Venant–Kirchhof hyperelastic model is used with $E\,$\,$=$\,$0.3\,MPa$ and $\nu = 0.3$. 
In our case, the Neo-Hookean formulation described in Section \ref{sec:solid} is employed.
In the context of rather small deformations, the two models have been reported to behave similarly \cite{TurekAnev}.
The ﬂuid is initially at rest and the tube is clamped at both ends.
On $\Gamma_{f,out}$, the pressure is set to zero, whereas $1.3 \times 10^3 Pa$ is prescribed on $\Gamma_{f,in}$ for $3\,ms$, before it is relaxed to zero, creating a pressure wave that travels along the tube.
We reproduced the case and solved it with a timestep of $\delta t = 0.05\,ms$, using first and second-order backward differentiation formulas for the fluid and solid time derivatives, respectively.
The employed timestep is halved compared to the original benchmark and was required in our case due to the more dissipative behaviour of our stabilized solid solver \cite{NEMER2021113923}.
The tolerance for assessing FSI subiteration convergence is set to $10^{-9}\,mm$.
Three different meshes have been tested to assess the convergence of the method, whose properties are summarized in Table \ref{tab:meshesPW3}.
As shown in Figure \ref{fig:PW3_definition}, a boundary layer is implemented in the fluid, to better capture the physics at the interface.
For this pressure wave benchmark, the accuracy of the results is usually assessed based on the radial and longitudinal displacements of a point situated at the middle of the tube, on the inner surface (see point A in Figure \ref{fig:PW3_definition}).
The obtained curves for the three meshes are displayed in Figure \ref{fig:PW3_results} along with reference curves taken from \cite{Eken, LOZOVSKIY, Ryzha}.
The quality of the fit is encouraging and comforts the results reported in the following.
Pressure contours are also reported in Figure \ref{fig:PW3_pressure_streamlines}, showing excellent agreement with the work of Lozovskiy et al. \cite{LOZOVSKIY}.

\begin{table}[h!]
\caption{Characteristics of the pressure wave meshes. $h_{iso}$ stands for the element size in the core of the fluid.}

	\centering
	\begin{tabularx}{\linewidth}{
	p{\dimexpr.1\linewidth-2\tabcolsep-1.3333\arrayrulewidth}
	p{\dimexpr.21\linewidth-2\tabcolsep-1.3333\arrayrulewidth}
    p{\dimexpr.13\linewidth-2\tabcolsep-1.3333\arrayrulewidth}
    p{\dimexpr.27\linewidth-2\tabcolsep-1.3333\arrayrulewidth}
    p{\dimexpr.27\linewidth-2\tabcolsep-1.3333\arrayrulewidth}
	}
	Mesh Id. & nb. layers in the solid & $h_{iso}$ (mm) & nb. elements in the fluid mesh & nb. elements in the solid mesh\\
	\Xhline{3\arrayrulewidth}
    C & 4 & 0.70 & 176k & 64k\\
    \hline
    M & 6 & 0.48 & 365k & 186k\\
    \hline
    F & 8 & 0.35 & 634k & 409k\\
    \Xhline{3\arrayrulewidth}

\end{tabularx}
\label{tab:meshesPW3}
\end{table}

\subsection{FSI in the proposed idealized aneurysm (R)}
\begin{figure}[h]
    \centering
    \begin{tikzpicture}
    \node at (0,0) {
    \begin{tabular}{c c}
        & \textit{rigid tissue modelling \hspace{4cm} FSI modelling}\\
        \toprule
        \rotatebox{90}{\hspace{65pt} \textit{R2}}
        &\includegraphics[width=0.67\linewidth, clip]{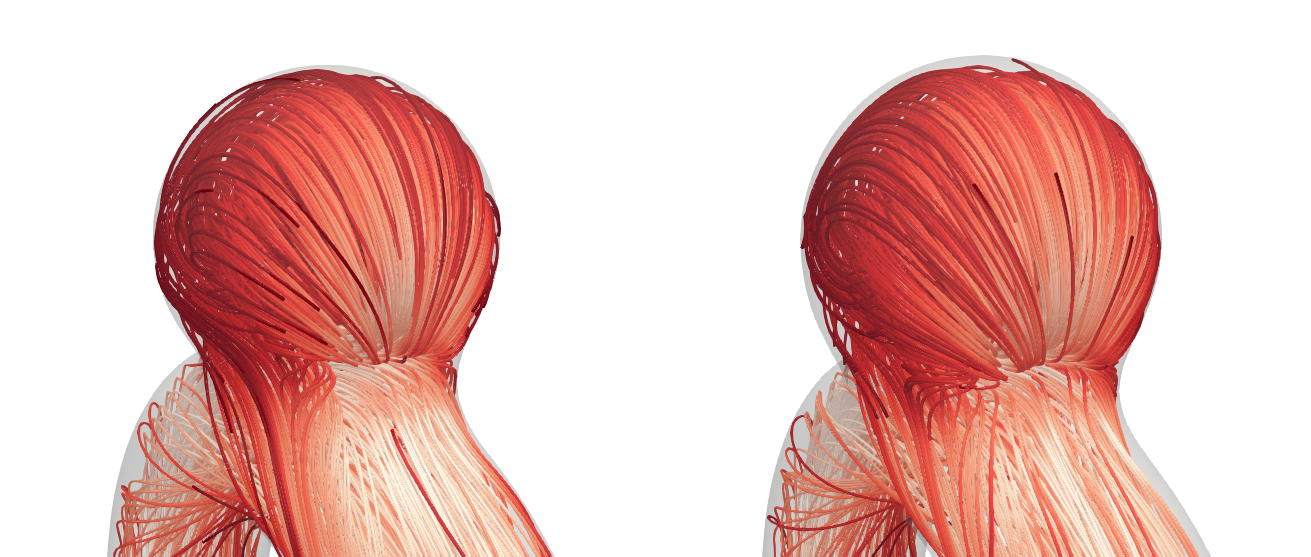}\\
        \toprule
        \rotatebox{90}{\hspace{65pt} \textit{S1}}
        &\includegraphics[width=0.67\linewidth, clip]{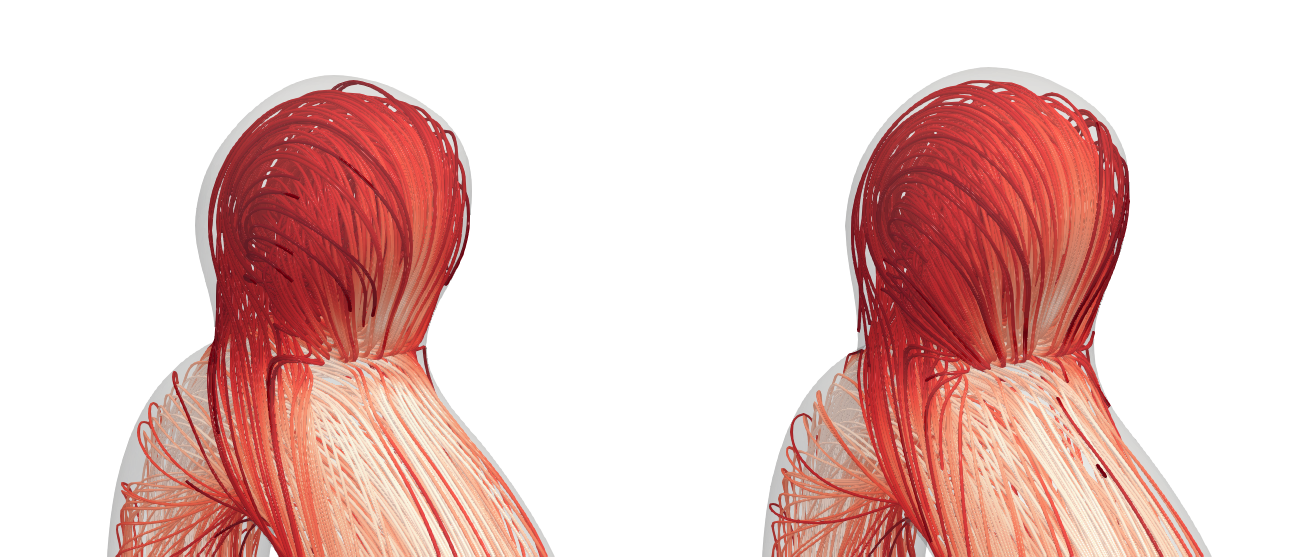}\\
        \toprule
        \rotatebox{90}{\hspace{65pt} \textit{S2}}
        &\includegraphics[width=0.67\linewidth, clip]{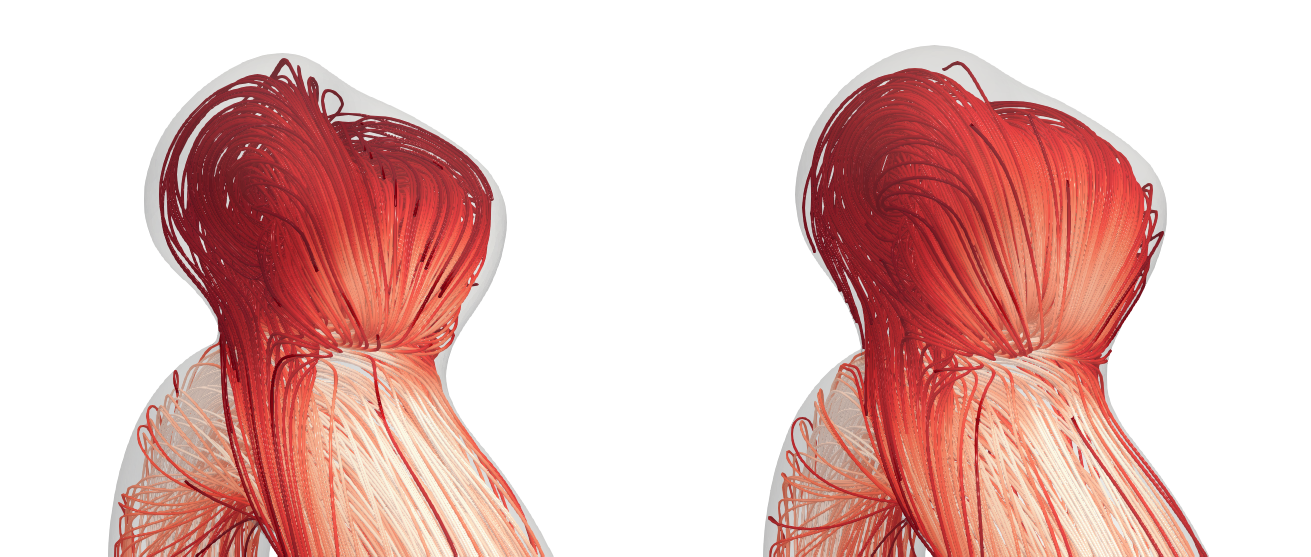}\\
        \toprule
        \rotatebox{90}{\hspace{65pt} \textit{S3}}
        &\includegraphics[width=0.67\linewidth, clip]{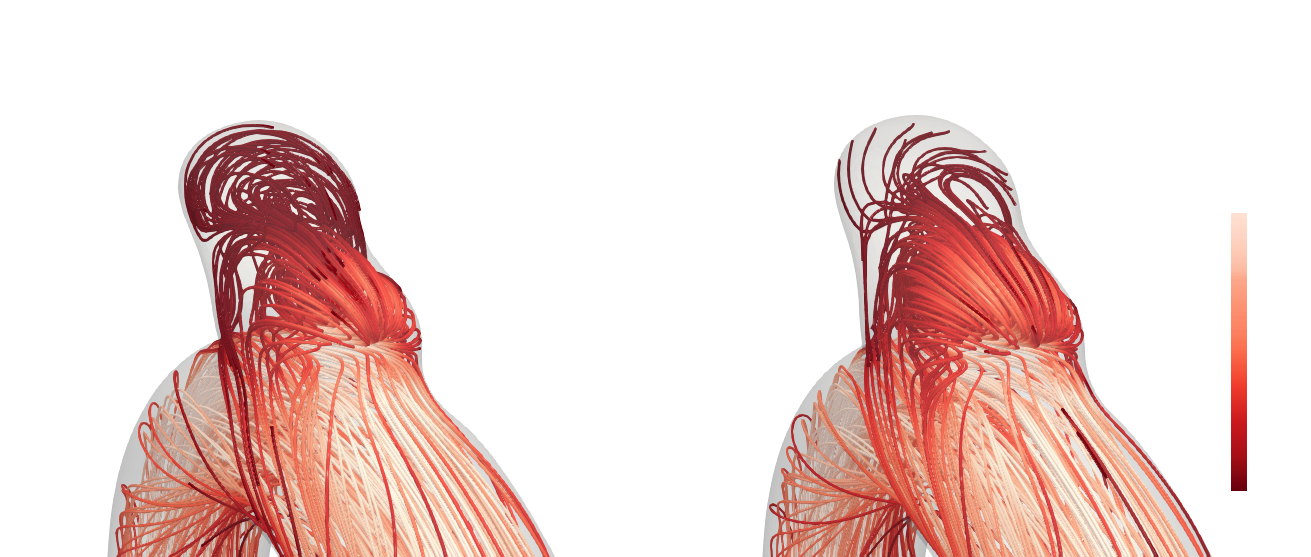}\\
        \toprule
    \end{tabular}
    };
    \node[rotate=90] at (6.5,-8.75) {\fontsize{9}{12}\selectfont Velocity (m/s)};
    \node at (6,-7.6) {1};
    \node at (6,-10.1) {0};
    \end{tikzpicture}
    \caption{Systolic flow lines for both rigid and compliant wall modellings.}
    \label{fig:streamlines}
\end{figure}

\begin{figure}[h]
    \centering
    \begin{tikzpicture}
    \node at (0,0) {
    \begin{tabular}{c c}
        & \textit{rigid tissue modelling \hspace{4cm} FSI modelling}\\
        \toprule
        \rotatebox{90}{\hspace{65pt} \textit{R2}}
        &\includegraphics[width=0.67\linewidth, clip]{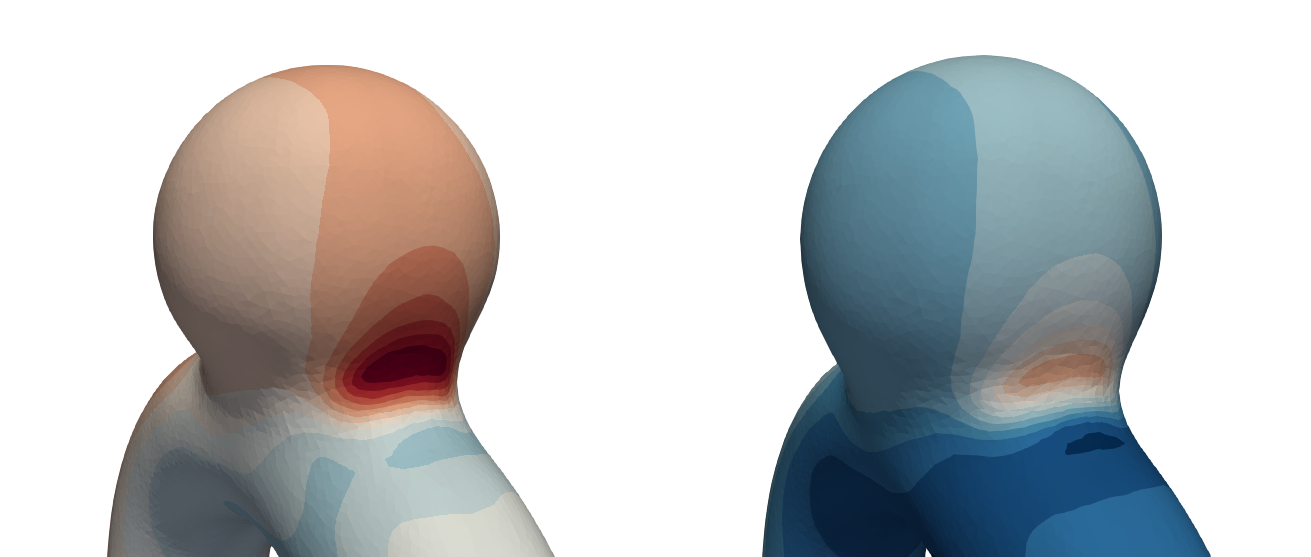}\\
        \toprule
        \rotatebox{90}{\hspace{65pt} \textit{S1}}
        &\includegraphics[width=0.67\linewidth, clip]{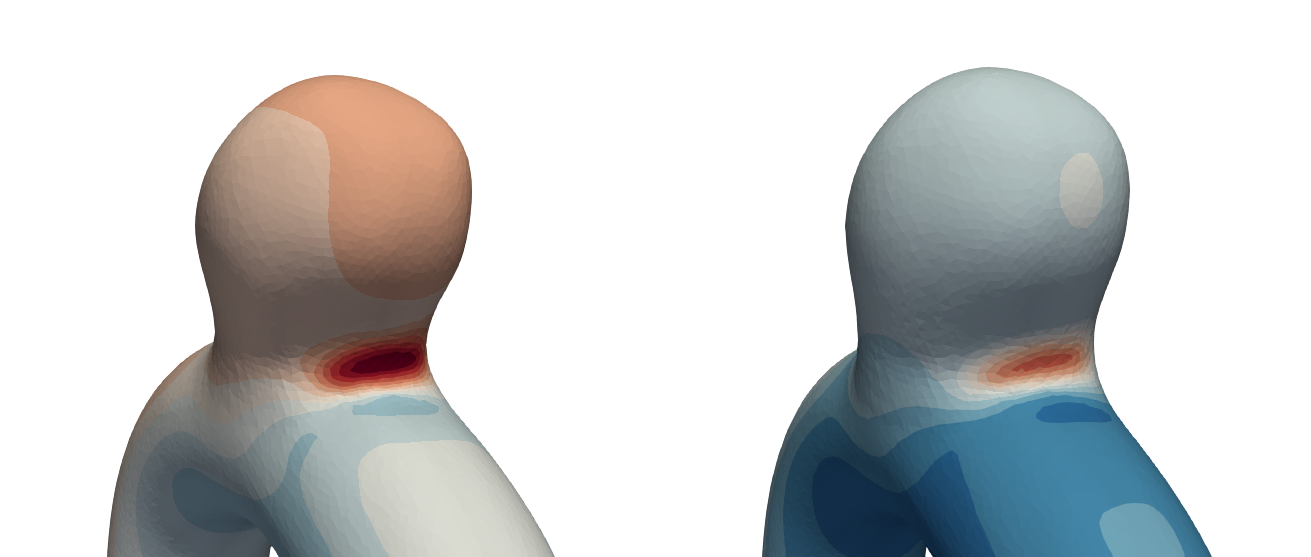}\\
        \toprule
        \rotatebox{90}{\hspace{65pt} \textit{S2}}
        &\includegraphics[width=0.67\linewidth, clip]{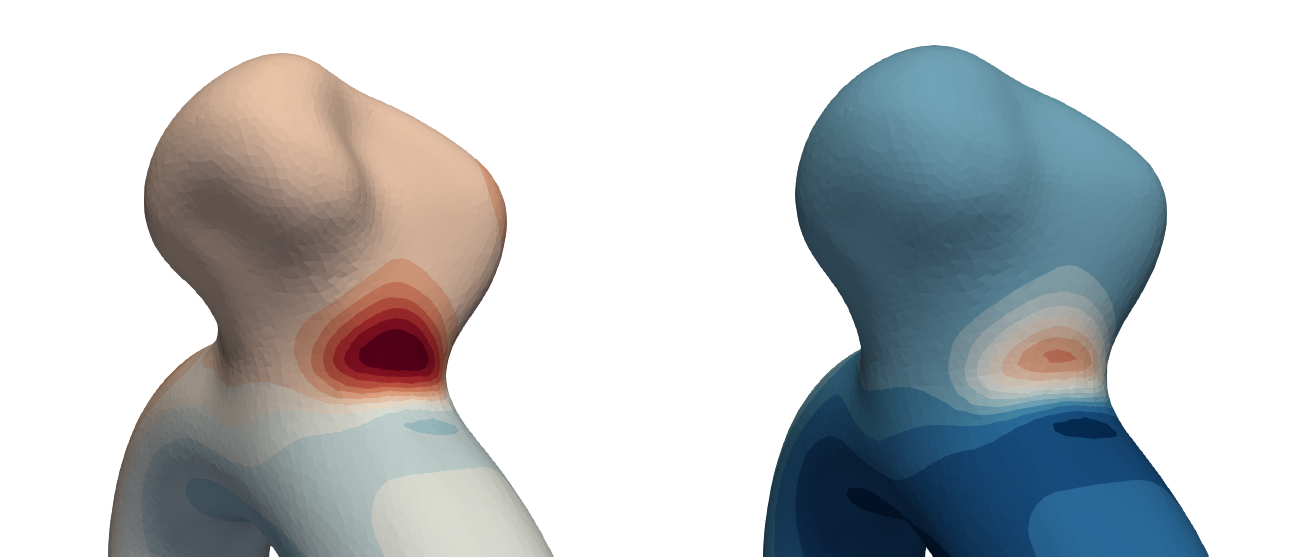}\\
        \toprule
        \rotatebox{90}{\hspace{65pt} \textit{S3}}
        &\includegraphics[width=0.67\linewidth, clip]{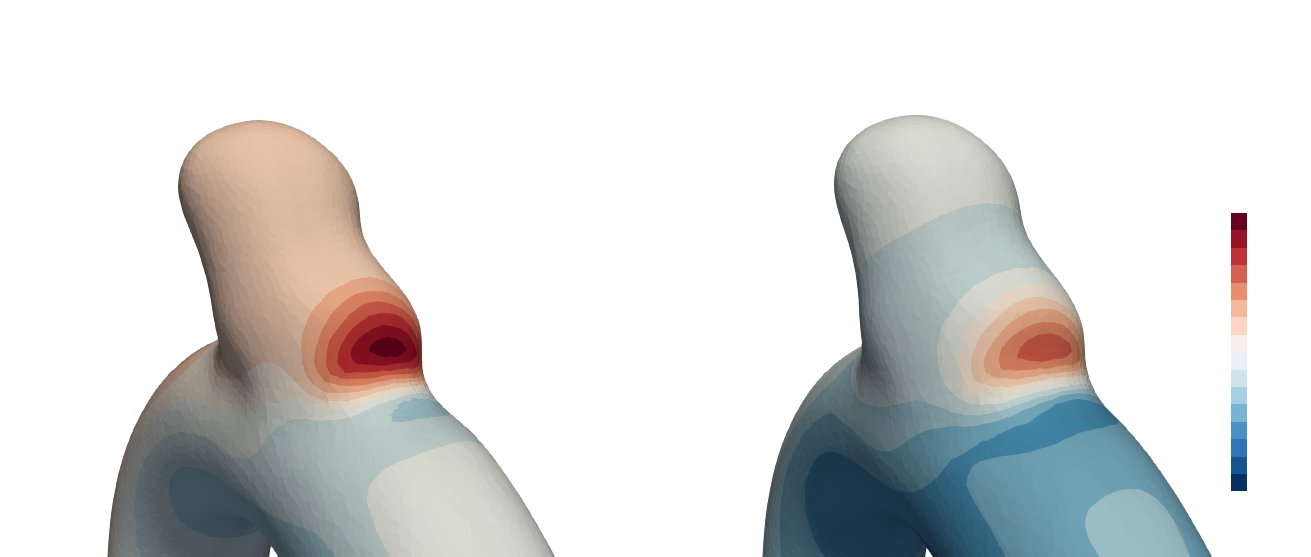}\\
        \toprule
    \end{tabular}
    };
    \node[rotate=90] at (6.5,-8.75) {\fontsize{9}{12}\selectfont Pressure (kPa)};
    \node at (6.1,-7.6) {5.7};
    \node at (6.1,-10.1) {4.6};
    \end{tikzpicture}
    \caption{Systolic pressure for both rigid and compliant wall modellings.}
    \label{fig:pressure}
\end{figure}

\begin{figure}[h!]
    \centering
    \begin{tabular}{c c}
        & \hspace{0.2cm} \textit{Systolic velocity iso-surface \hspace{0.3cm} Systolic fluid domain extension}\\
        \toprule
        \rotatebox{90}{\hspace{34pt} \textit{R2}}
        &\includegraphics[width=0.43\linewidth, clip]{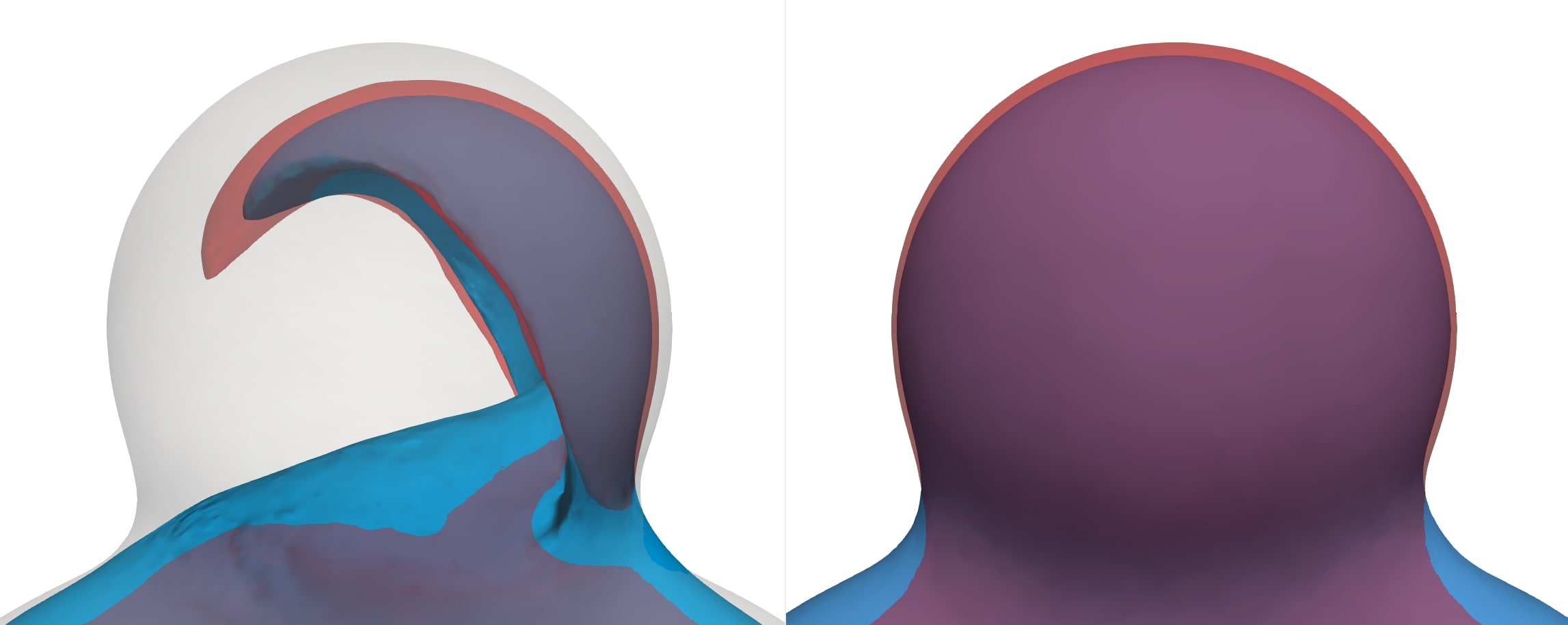}\\
        \toprule
        \rotatebox{90}{\hspace{34pt} \textit{S1}}
        &\includegraphics[width=0.43\linewidth, clip]{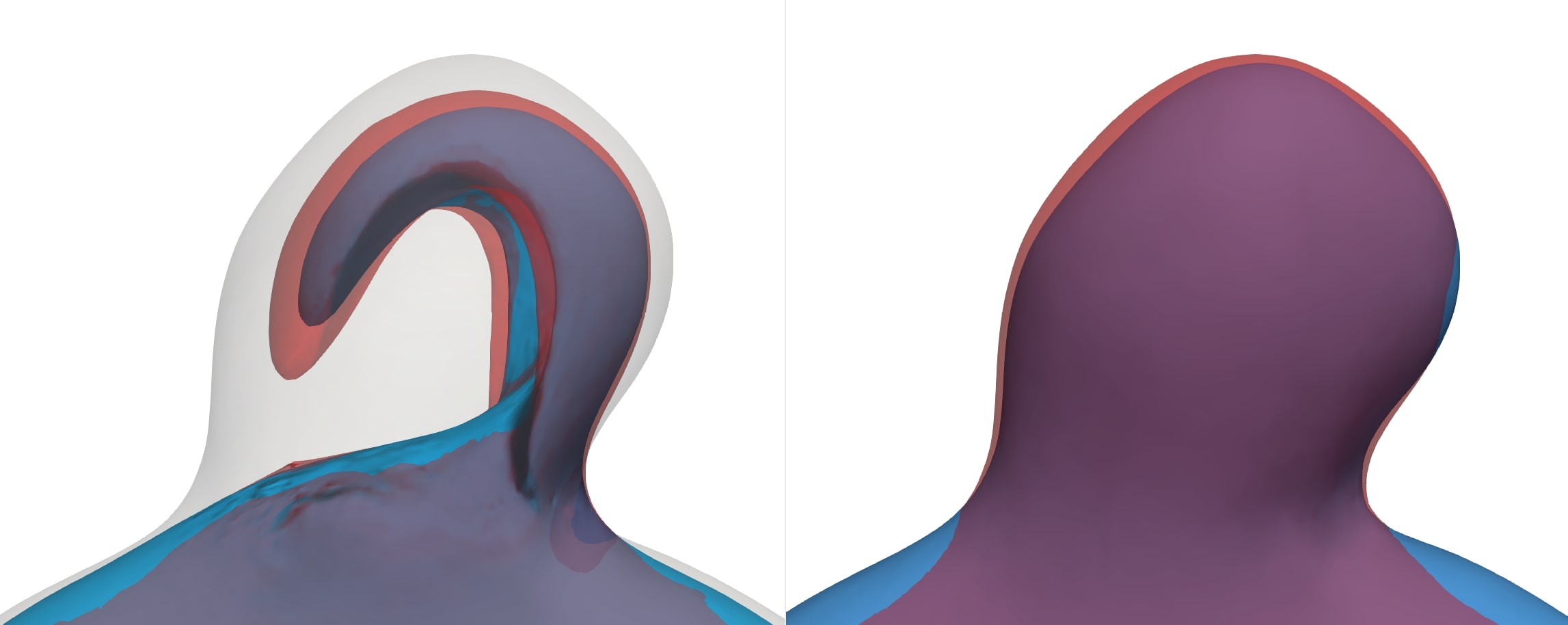}\\
        \toprule
        \rotatebox{90}{\hspace{34pt} \textit{S2}}
        &\includegraphics[width=0.43\linewidth, clip]{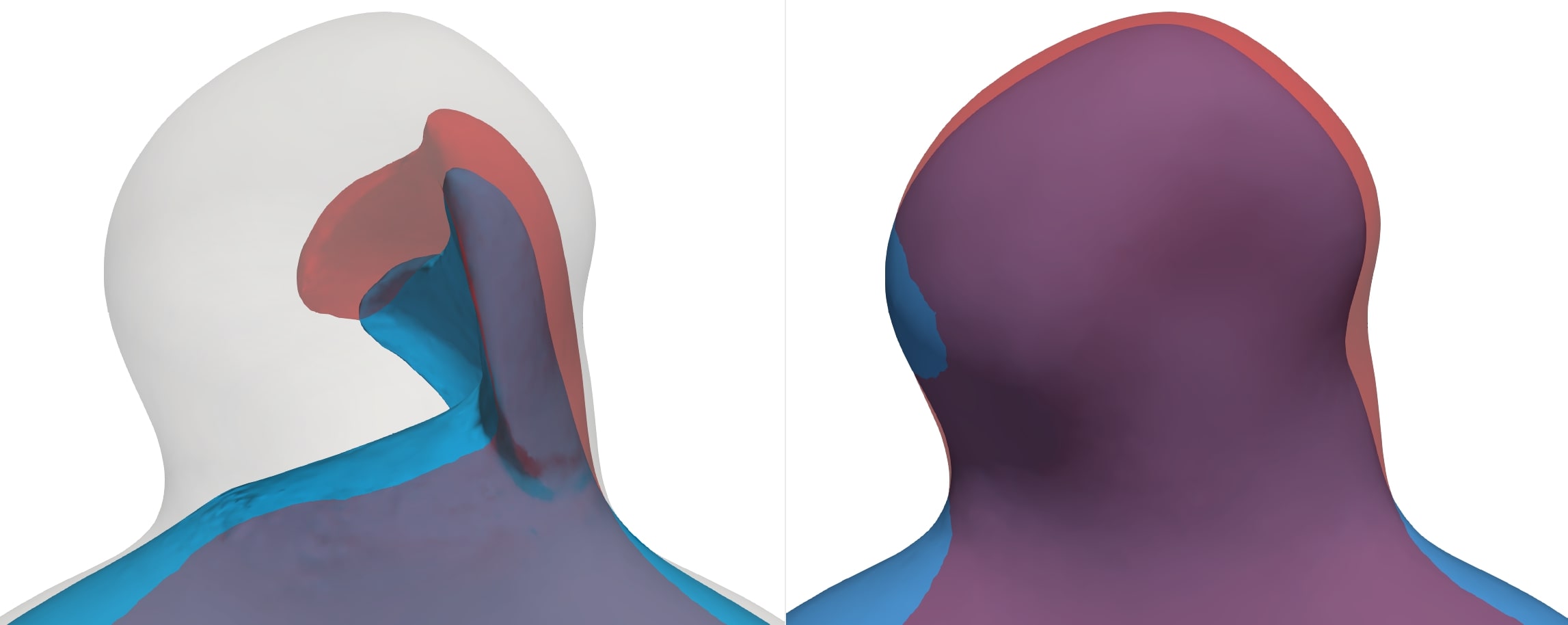}\\
        \toprule
        \rotatebox{90}{\hspace{34pt} \textit{S3}}
        &\includegraphics[width=0.43\linewidth, clip]{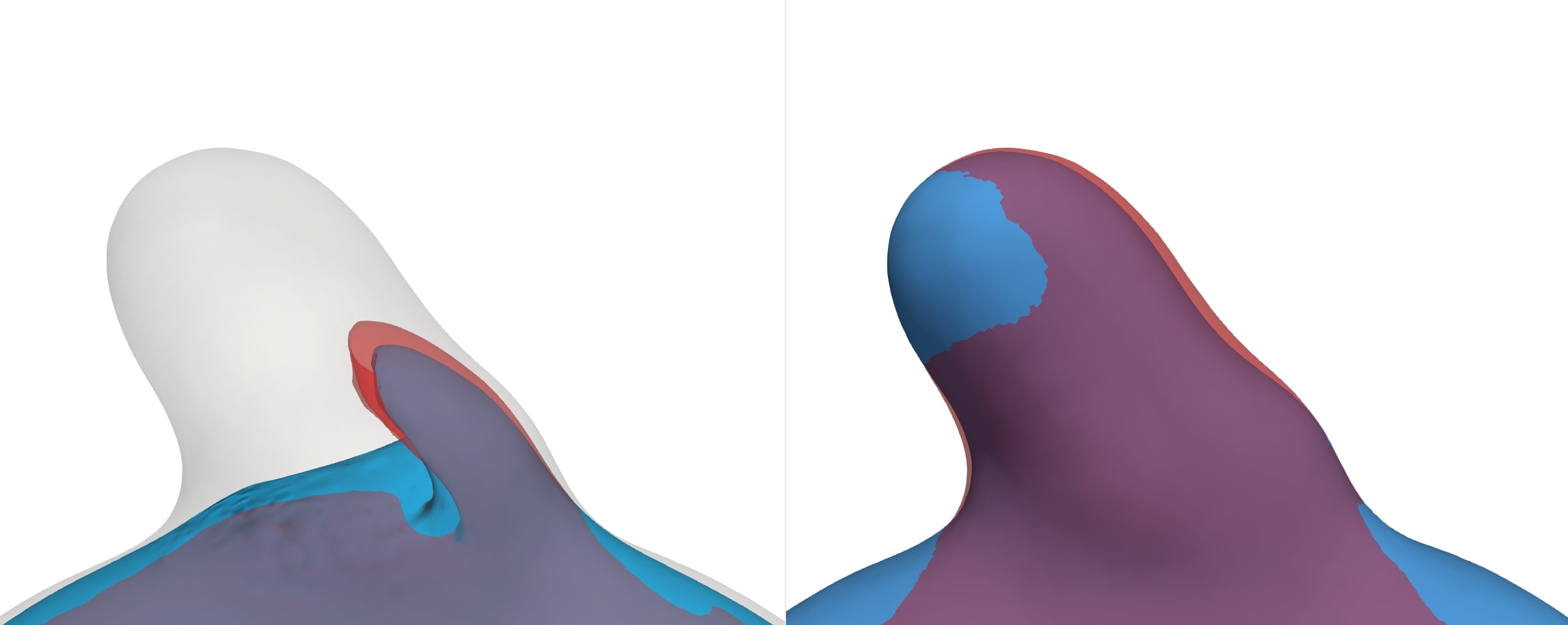}\\
        \toprule
    \end{tabular}
    \caption{Arbitrary systolic velocity iso-surface (left) and fluid domain extension (right), for both rigid (blue) and compliant (red) tissue modellings. Geometries are viewed from the side (along the $z$ axis).}
    \label{fig:isosurfaceANDextension}
\end{figure}

Before moving to complex bulge shapes, the impact of FSI modelling on the spherical aneurysm is assessed in this section.
To better visualize the 3D flow patterns involved, systolic velocity streamlines are given in Figure \ref{fig:streamlines}, both in the rigid and compliant configurations.
Similarly, the pressure field is shown in Figure \ref{fig:pressure}.
To easily understand the impact of the modelling type on the aneurysmal flow profile and on the domain's expansion, Figure \ref{fig:isosurfaceANDextension} gives a systolic velocity iso-surface along with an overlay of the diastolic and systolic fluid domain shapes.
In that latter, it can be observed that modelling tissue compliance creates a deeper swirl of the flow inside the bulge, due to an opening of the neck mostly in the $z$-direction.
Apart from this shift, the general flow pattern is barely affected by the compliant tissue modelling.
In both cases, the flow separates at the neck, with a small fraction entering the bulge, impinging on the wall, and resulting in high WSS values (TAWSS rise up to around $200\,dyn.cm^{-2}$ in the FSI case).
The membrane's movement is dominantly pressure-driven, with almost no tilting of the aneurysm in response to the inflow jet at the neck, agreeing with previous studies \cite{Tor+08}.
Contrary to velocity, the pressure field is consequently affected by the modelling choice.
Indeed, the geometry compliance allows for delayed flow variations between the inlet and outlet, which directly affects the prescribed boundary conditions \eqref{equation:BC}.
For instance, the system is storing volume at systole, resulting in a lower outflow pressure.
It can also be observed that the compliant modelling relaxes the pressure peak at the impingement area.

The impact of the modelling choice on the quantities of interest (see Section \ref{sec:qoi}) is reported in Figure \ref{fig:WSS} and \ref{fig:OSI}.
As the bulge opens under fluid stress, the wall inclination at the impingement area increases.
As a result, this impact spot is shifted up, explaining the small region of higher TAWSS (in blue) for the FSI modelling in Figure \ref{fig:WSS}.
Apart from this specific spot, compliant modelling generally lowers the WSS peak values in the bulge as reported in Table \ref{tab:results}, confirming trends of previous studies \cite{Baz+10, Tor+09}. 
However, as flow penetrates farther into the aneurysm dome, WSS at the fundus increases and the high OSI values are shifted towards the back of the bulge.
For R2, fundus TAWSS varies by 15.7\% with respect to the rigid configuration (see Table \ref{tab:results}).
All the reported effects are naturally amplified by lower stiffness values, which result in higher bulge deformation.
For the explored configurations, observed volume variations and displacements lie in the range of previously reported values \cite{Van+15, Sta+21}, with R3 being in the extreme upper spectrum.

Even though, the variables of interest change quantitatively, general distributions remain very similar, even when pushing the material properties to low values (R3).
The embedded symmetry of the reference case (R) naturally reflects on flow patterns resulting in smaller compliance-driven changes compared to patient-specific cases \cite{Baz+10, VoS+16}.

\begin{figure}[h!]
    \centering
    \begin{tikzpicture}[scale=2.2]
    \node at (0,0) {
    {\includegraphics[width=1\linewidth]{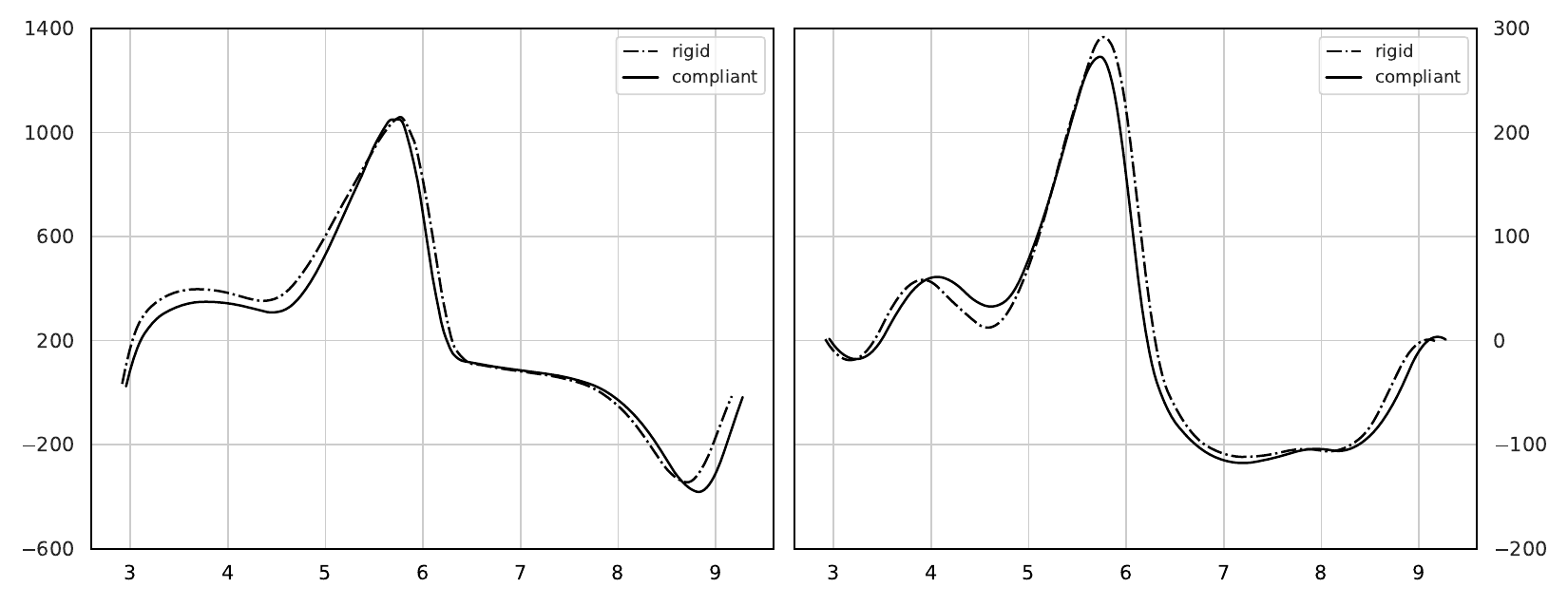}}
    };
    \node at (2.77,0.47) {
    {\includegraphics[width=0.18\linewidth]{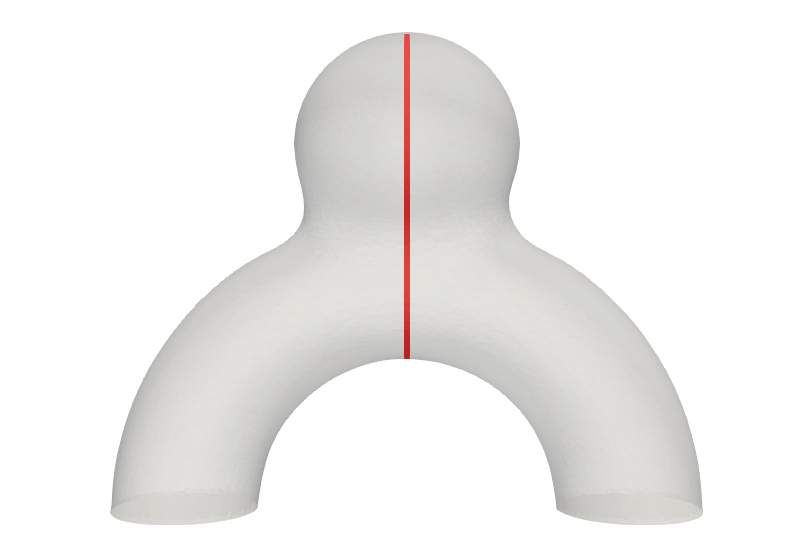}}
    };
    \node at (-0.86,0.47) {
    {\includegraphics[width=0.18\linewidth]{plot_over_line.png}}
    };
    \node[rotate=90] at (4.05,0) {$\bm{v}_y$ (mm/s)};
    \node[rotate=90] at (-4.05,0) {$\bm{v}_x$ (mm/s)};
    \node at (-1.8,-1.6) {$y$ coordinate (mm)};
    \node at (2,-1.6) {$y$ coordinate (mm)};
    \end{tikzpicture}
    \caption{Reference systolic velocity profiles ($t=1.08\,s$) along the $y$ axis (see. Figure \ref{fig:AnXplore_reference_definition}) with rigid and compliant (R2) walls.}
    \label{fig:R2_ref_curve}
\end{figure}

Similar to the pressure wave benchmark, convergence has been assessed for R2 and velocity profiles are reported in the Supplementary materials for several mesh refinements and time steps.
Using the retained mesh (described in Section \ref{sec:meshing}) and a timestep of $\Delta t = 5\,ms$, reference validation curves are proposed in Figure \ref{fig:R2_ref_curve}.
These plots provide the systolic velocity ($t = 1.08\,s$) along the $y$ axis, which vertically goes through the aneurysm bulge.
The plots are given for both rigid and compliant models to ease a step-by-step reproduction of the presented results.

\clearpage

\begin{table}[h!]
\caption{Comparison of FSI and rigid models through quantities of interest. All WSS-based relative differences are computed with the compliant model as a reference. The fundus location is identified on the rigid configuration as the point featuring the lowest TAWSS.}

	\centering
	\begin{tabularx}{\linewidth}{
	p{\dimexpr.13\linewidth-2\tabcolsep-1.3333\arrayrulewidth}
	p{\dimexpr.2\linewidth-2\tabcolsep-1.3333\arrayrulewidth}
    p{\dimexpr.21\linewidth-2\tabcolsep-1.3333\arrayrulewidth}
    p{\dimexpr.25\linewidth-2\tabcolsep-1.3333\arrayrulewidth}
    p{\dimexpr.21\linewidth-2\tabcolsep-1.3333\arrayrulewidth}
	}
	Case Id. & Max volume variation [\%] & Max displacement of $\Gamma_{FSI}$ (mm) &
	Systolic WSS drop at rigid impingement spot [\%] & Funuds TAWSS increase [\%]\\
	\Xhline{3\arrayrulewidth}
    R1 & 6.8 & 0.20 & 4.1 & 8.9\\
    \hline
    R2 & 11.5 & 0.26 & 9.8 & 15.7\\
    \hline
    R3 & 20.9 & 0.33 & 22.7 & 22.7\\
    \Xhline{3\arrayrulewidth}

    S1 & 12.4 & 0.23 & 2.5 & 22.8\\
    \hline
    S2 & 16.7 & 0.45 & 16.3 & 79.0\\
    \hline
    S3 & 12.4 & 0.21 & -0.8 & 47.6\\
	\Xhline{3\arrayrulewidth}
\end{tabularx}

\label{tab:results}
\end{table}

\subsection{Exploring different bulge shapes (S)}

\begin{figure}[h!]
    \centering
    \begin{tikzpicture}
    \node at (0,0) {
    \begin{tabular}{c c}
        \toprule
        \rotatebox{90}{\hspace{60pt} \textit{R2}}
        &\includegraphics[width=0.9\linewidth, clip]{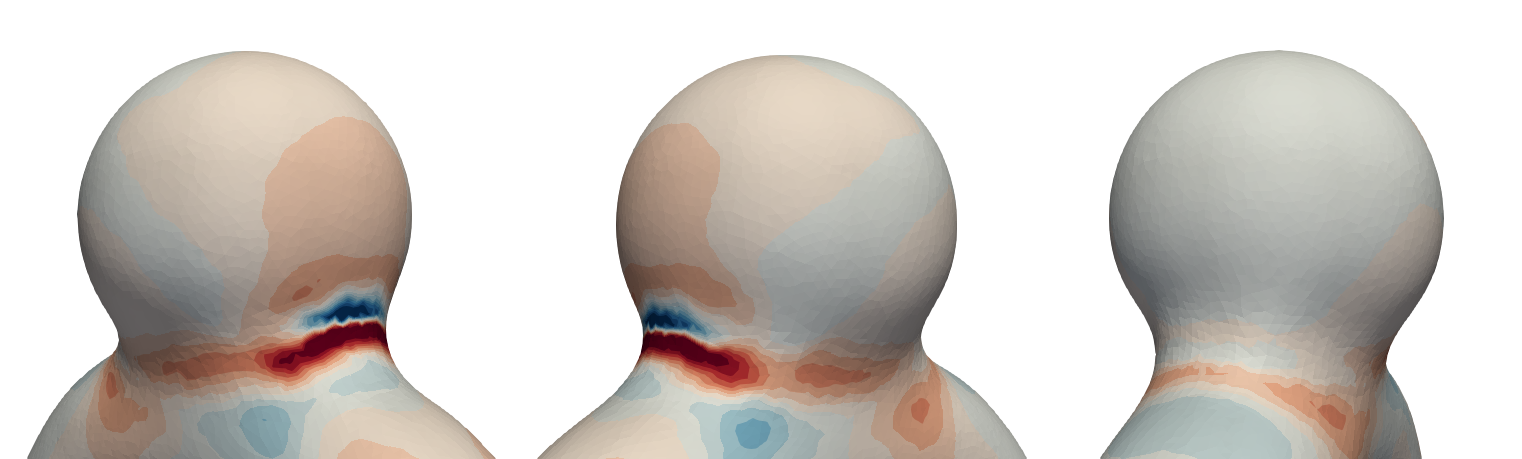}\\
        \noalign{\vskip 2mm} 
        \toprule
        \rotatebox{90}{\hspace{60pt} \textit{S1}}
        &\includegraphics[width=0.9\linewidth, clip]{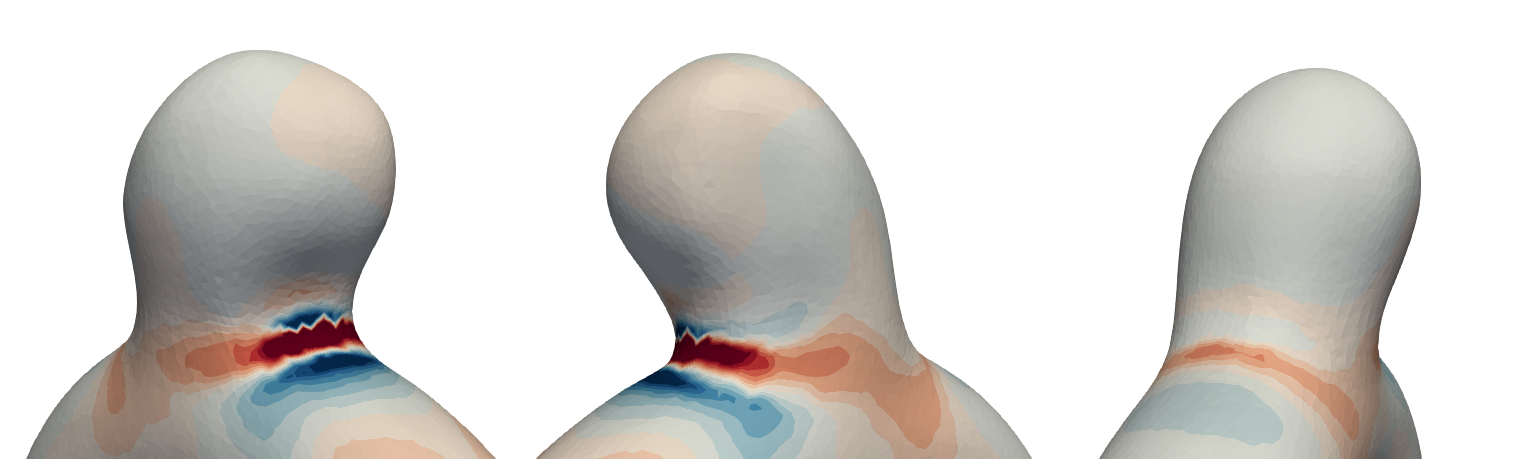}\\
        \noalign{\vskip 2mm} 
        \toprule
        \rotatebox{90}{\hspace{60pt} \textit{S2}}
        &\includegraphics[width=0.9\linewidth, clip]{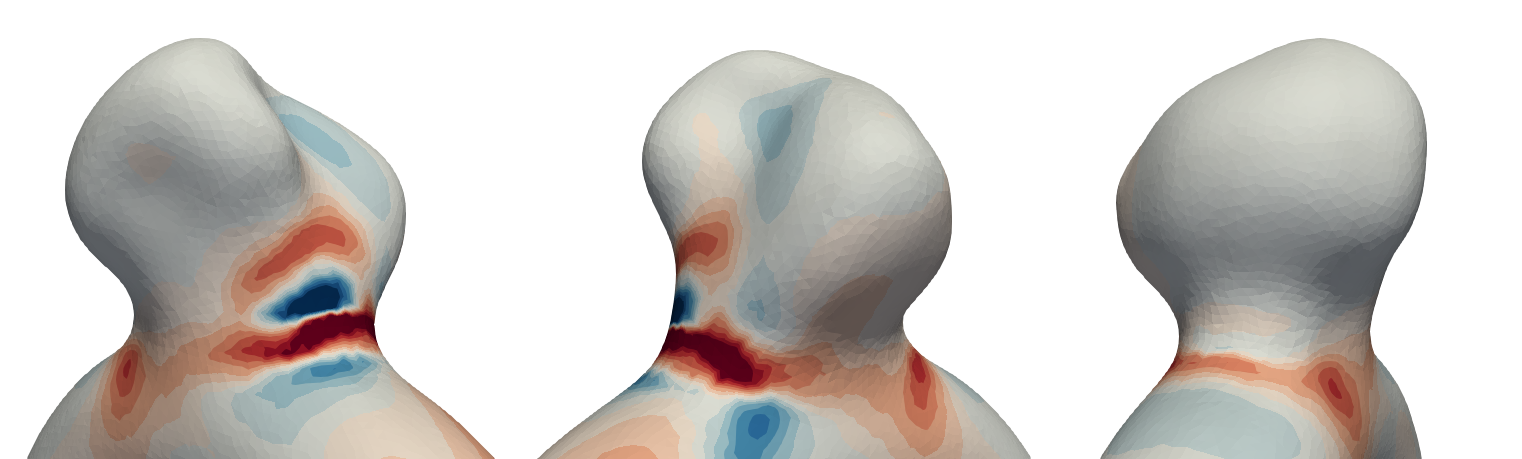}\\
        \noalign{\vskip 2mm} 
        \toprule
        \rotatebox{90}{\hspace{60pt} \textit{S3}}
        &\includegraphics[width=0.9\linewidth, clip]{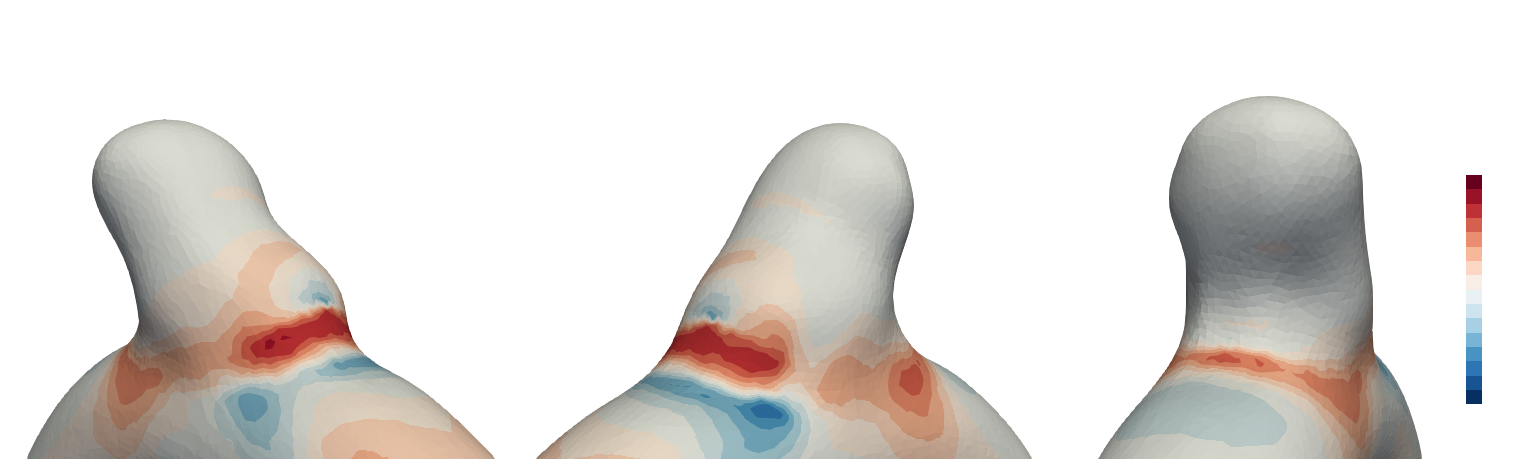}\\
        \noalign{\vskip 2mm} 
        \toprule
    \end{tabular}
    };
    \node[rotate=90] at (8.4,-8.5) {\fontsize{9}{12}\selectfont $\Delta$ TAWSS};
    \node[rotate=90] at (8.75,-8.5) {\fontsize{9}{12}\selectfont ($dyn.cm^2$)};
    \node at (7.95,-6.9) {+40};
    \node at (7.95,-10) {-40};
    
    \end{tikzpicture}
    \caption{TAWSS difference between the two wall modelling assumptions (rigid - compliant).}
    \label{fig:WSS}
\end{figure}

\begin{figure}[hb!]
    \centering
    \begin{tikzpicture}
    \node at (0,0) {
    \begin{tabular}{c c}
        \toprule
        \rotatebox{90}{\hspace{60pt} \textit{R2}}
        &\includegraphics[width=0.9\linewidth, clip]{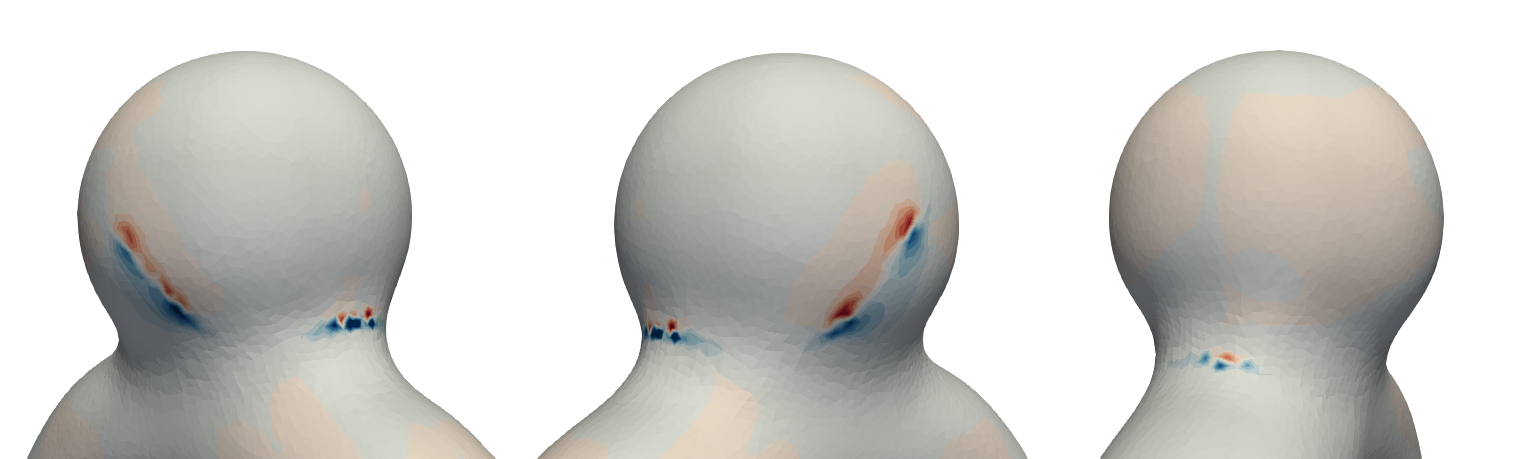}\\
        \noalign{\vskip 2mm} 
        \toprule
        \rotatebox{90}{\hspace{60pt} \textit{S1}}
        &\includegraphics[width=0.9\linewidth, clip]{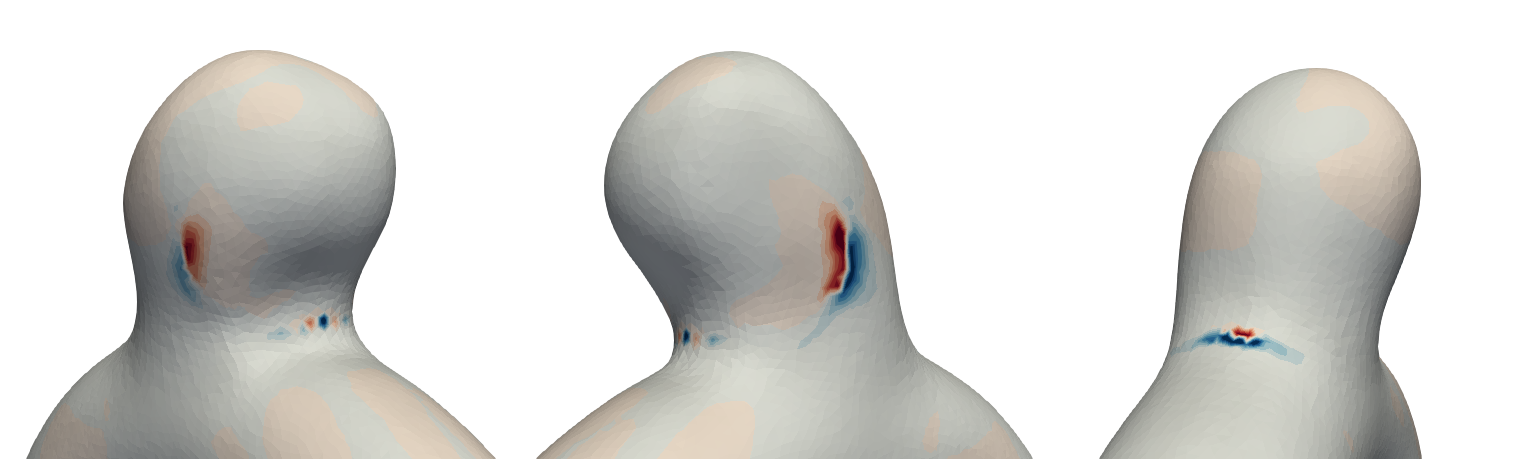}\\
        \noalign{\vskip 2mm} 
        \toprule
        \rotatebox{90}{\hspace{60pt} \textit{S2}}
        &\includegraphics[width=0.9\linewidth, clip]{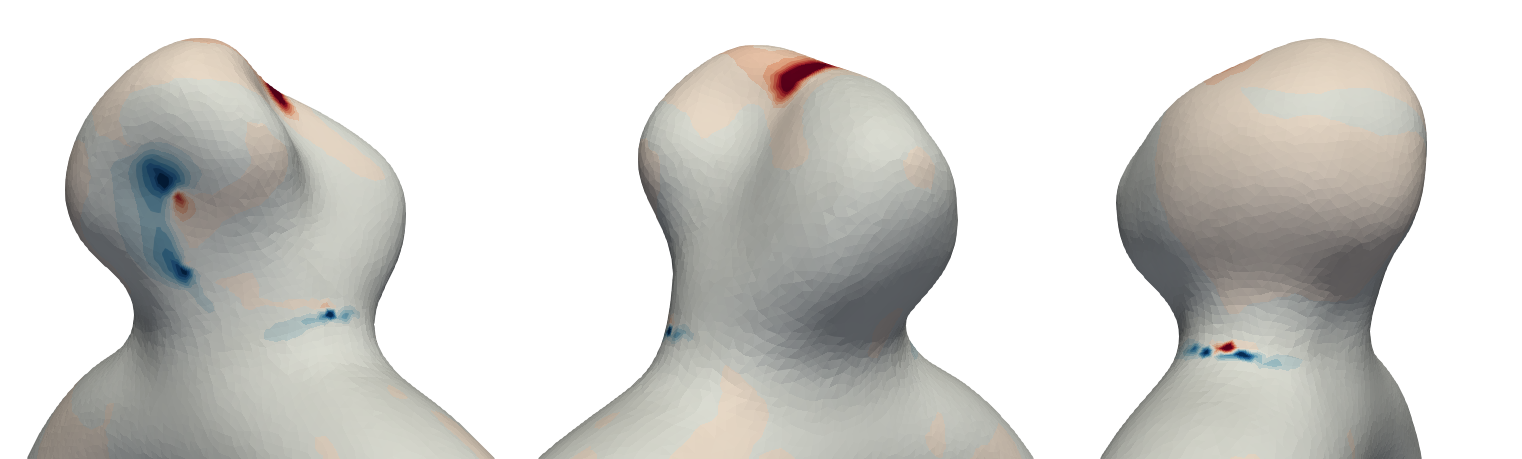}\\
        \noalign{\vskip 2mm} 
        \toprule
        \rotatebox{90}{\hspace{60pt} \textit{S3}}
        &\includegraphics[width=0.9\linewidth, clip]{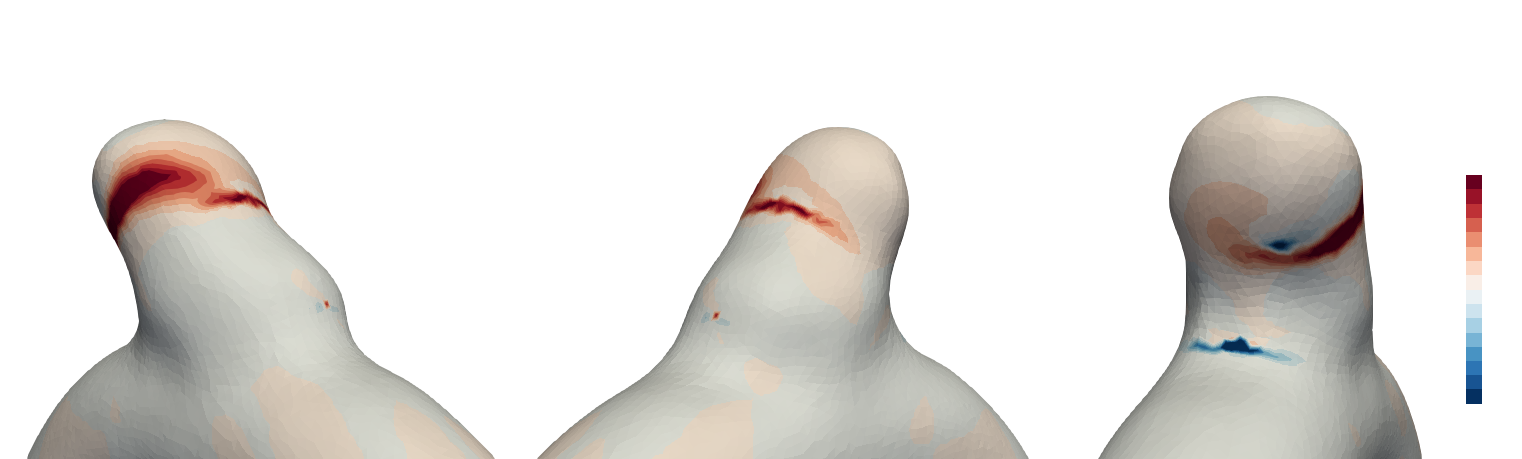}\\
        \noalign{\vskip 2mm} 
        \toprule
    \end{tabular}
    };
    \node[rotate=90] at (8.4,-8.55) {\fontsize{9}{12}\selectfont $\Delta$ OSI};
    \node at (7.95,-6.9) {+0.2};
    \node at (7.95,-10) {-0.2};
    \end{tikzpicture}
    \caption{OSI difference between the two wall modelling assumptions (rigid - compliant).}
    \label{fig:OSI} 
\end{figure}

To explore implications beyond a fully spherical bulge, the specific shapes S1-3 are thoroughly examined in Figures \ref{fig:streamlines}-\ref{fig:OSI}, similar to R2.
Please note that, by construction, all the cases have the same neck geometry, which allows drawing the focus solely on the bulge's topology.

S1-3 cases share some general flow aspects with the reference spherical geometry.
Using FSI modelling, the flow swirls deeper and the impingement area shifts upwards in the bulge, resulting in an overall lower TAWSS.
Although inspired by a real aneurysm shape, S1 does not deviate significantly from R2, even when looking at the details.
S2 and S3, per contrary, have very different flow mechanisms, leading to large deviations between rigid and compliant tissue modelling.

First, the S2 geometry expands mainly in the impingement area (as displayed in Figure \ref{fig:isosurfaceANDextension}), altering the inflow jet and reducing by 16.3\% the WSS at the impact region predicted by the rigid model.
As shown in Figure \ref{fig:streamlines}, the initially split aneurysmal flow (rigid configuration) is merged in one general swirl for the compliant configuration due to the outward bulging of the solid in the impingement area.
As a consequence, the dome's exposure to the flow increases, thereby locally rising TAWSS by 79\%.
The overall flow change is also illustrated by shifted OSI patterns, translating the alteration of the swirls when modelling the compliant tissue behaviour.

The last investigated case, S3, also shows a noticeable change in flow pattern when employing compliant modelling.
Contrary to S1-2, the flow does not penetrate easily the dome due to the orientation of the bulge with respect to the parent vessel.
This results in a secondary recirculation at the dome in the rigid case, which is commonly observed in high aspect-ration IAs. 
This slow recirculation is significantly altered when compliant modelling is employed, due to its little inertia.
The rotation of the flow is indeed modified by the expansion and contraction of the wall as shown in Figure \ref{fig:streamlines}.
Even though TAWSS is not altered significantly, systolic WSS at the fundus changes by 47.6\%.
The OSI pattern is also largely impacted, as it is very sensitive to flow orientation (see Figure \ref{fig:OSI}). 
Overall, the reported results show how different bulge shapes may demonstrate various responses to compliant tissue modelling.

\section{Discussion}
\subsection{Impact of the wall modelling}

Initiation and growth of IAs are sparked by abnormal flow conditions through different cascades of biological reactions \cite{MAl99, Fro+04}.
This tissue remodelling has been shown to be strongly correlated with haemodynamic metrics such as the ones investigated in this work \cite{Men+14}.
The most commonly studied remains the WSS, for which both high and low values are considered a threat to IA's stability over time.
If high values emphasize areas of fluid stress concentration, the lower ones are associated with blood stagnation, which can trigger an inflammatory response of the walls and atherosclerosis development \cite{HMY06}.
The OSI complements the description at the wall's vicinity by giving insights on flow orientation changes over a cardiac cycle.
The fact that these metrics can change significantly depending on the employed wall modelling suggests that rigid CFD results may yield inaccurate risk estimation.
Compliant tissue modelling apparently affects WSS extreme values typically by shrinking the range.
In the second specific case (S2), the irregular bulge topology features a weak spot where the largest deformation occurs.
Because this structural weakness is located in the impingement area where most of the flow enters the bulge, the overall pattern is consequently modified and peak systole WSS values decrease by more than 15\%.
Concerning the low values, this case has also revealed a 79\% TAWSS increase at the dome due to the slight inflation of the bulge that renders the fundus more accessible.
Such slow-flow regions are particularly sensitive to wall compliance, as seen in the OSI patterns of S2 and S3.
For the latter case, the general haemodynamics have substantially changed at the dome, even leading to local flow inversions.
OSI has drastically increased there as the wall-induced flow pattern is more prone to changes of orientation over a cardiac cycle.
This information appears essential in terms of thrombus formation risk assessment, as some indicators are classically built based on OSI values \cite{Gut+19}.
Even though not studied here, the flow residence time is also commonly employed as an indicator of IA stability and to predict potential thrombus formation for numerically evaluating the outcome of flow-diverters \cite{Bep+20}.
Such a flow metric will also undoubtedly reveal a different behaviour using the compliant modelling and should be explored in future work to reveal the impact on different aneurysms. 
All in all, considering that bulge geometries with daughter sacs or intricate dome topologies are very common and often feature zones of blood stagnation, we deem crucial to employ a compliant wall modelling to retrieve accurate slow flow patterns.

While S2 and S3 feature some noticeable differences in terms of classical haemodynamic risk metrics, results have shown that S1's haemodynamics are barely altered, certainly because of the case's regular shape and the bulge alignment with the impinging flow.
The diverse responses towards wall modelling observed in this study lead to the hypothesis that compliance-related effects are strongly shape-dependent.
This motivates our sensitivity study along with future research effort to assess the generalizability of the drawn conclusions onto large-scale patient cohorts and help identify critical cases, for which compliant modelling would benefit the most.
The three specific cases reported in this work pave the way toward this improved comprehension.
Contrary to other patient-specific data inputs, the lumen's topology is systemically acquired in clinical routine when detecting an aneurysm, already rendering such studies feasible.
However, as our results suggest, particular attention has to be dedicated to the segmentation quality of magnetic resonance angiography along with the level of smoothing employed when generating the computational domains \cite{Par+23}, as topological details can yield large deviations when compliant modelling is employed.

Lastly, an important aspect regarding outflow boundary conditions has to be stressed.
It is interesting to note that almost all previous publications reporting FSI simulations of IAs investigated bifurcation cases \cite{Tor+08, Baz+10, Val+13, VoS+16, Isa+08}.
In this study, a single outlet branch has been considered for the sake of simplicity, thus circumventing the treatment of the outflow split.
If outflow boundary conditions have already proven to be a sensitive aspect of CFD simulations \cite{Ram+13}, it is even more for FSI.
Indeed, several outflow branches featuring different radii will inevitably expand differently under internal pressure, thus altering the hydraulic resistance of the outflow paths and modifying the flow split.
This change surely impacts the general flow pattern consequently.
This observation suggests that the reported differences between rigid-wall and FSI simulations could be amplified when considering intricate patient-specific geometries, which almost always feature vascular branches in the vicinity of the aneurysm.

\subsection{Limitations and perspectives}

The geometries investigated in this work remain idealized and, although efforts have been made to tailor the aneurysm bulge with different shapes, a gap with patient-specific cases remains.
The drawn conclusions give a grasp on the FSI modelling relevance but are not to be directly transposed to clinical applications.
However, validating the highlighted trends on large-scale patient cohorts should stand as a future goal for the community, as previous studies remained limited to a few investigated cases.
Following the same line, boundary conditions could be expanded to reach high-fidelity modelling of realistic aneurysms.
Outflow pressure could be computed through a Windkessel model, and the solid structure pre-loaded as done in \cite{Baz+10}.
These details do not stand as a major technical burden but their relevance in such a simplified geometry remains unclear, motivating our inclination to simplicity.
Finally, realistic tissue material behaviour and thickness should be prescribed to achieve accurate conclusions.
Nevertheless, considering the lack of patient-specific data and of general guidelines regarding pathological tissue modelling, we believe that this belongs to future work.
Progress in imaging technology will give more information about these missing parameters, allowing accurate modelling tools to prove their efficiency.
If locally varying wall properties can be assessed \textit{in-vivo}, FSI simulations will appear even more relevant.
As local weaknesses are prone to result in larger deformations and constitute critical concentration points that jeopardize IAs' stability, FSI will surely contribute to high-fidelity risk assessment tools.
We hope that future work will provide insights into a categorization of IAs based on their sensitivity to FSI modelling.
While only a few shapes have been investigated as part of this study, the versatility of the proposed case allows us to explore a large manifold of realistic aneurysms in a controlled environment.
Underlining aneurysm phenotypes that mostly benefit from the compliant modelling of arterial tissue stands as one of our future research goals.

\section{Conclusion}

This work introduced a novel idealized sidewall aneurysm geometry for assessing the relevance of FSI modelling in various configurations.
Different tissue stiffness values as well as four aneurysm bulge shapes have been investigated, keeping all the other simulation parameters unchanged.
Well-known haemodynamic metrics such as WSS and OSI have been computed both using rigid and compliant wall modelling, revealing significant flow changes linked with the aneurysm topology.
Bulges featuring blebs in jet impingement areas, as well as aneurysms with slow recirculating flow regions have been emphasized as being more sensitive to FSI modelling.
The haemodynamic differences highlighted in this work suggest that modelling compliant vessels may be crucial for some IAs in order to predict their future growth and risk of rupture.
Differences between rigid and compliant modelling could even be amplified using patient-specific pathological tissue data, although literature does not provide insights yet.
Progress in medical imaging along with large-scale studies will certainly help to stress the limits of the widely-employed rigid wall assumption.
In all scenarios, FSI models like the one introduced in this work have to be developed to improve the comprehension of IAs and to assess the relevance of wall tissue modelling assumptions.

\section*{Acknowledgement}
This project has received funding from the European Research Council (ERC) under the European Union's Horizon Research and Innovation program (grant No. 101045042). We would like to thank the Neuro-Interventional department of the Nice University Hospital for their availability and support regarding medical aspects of this work.

\section*{Supplementary materials}

\begin{figure}[h!]
    \centering
    \begin{tikzpicture}[scale=2.2]
    \node at (0,0) {
    {\includegraphics[width=1\linewidth]{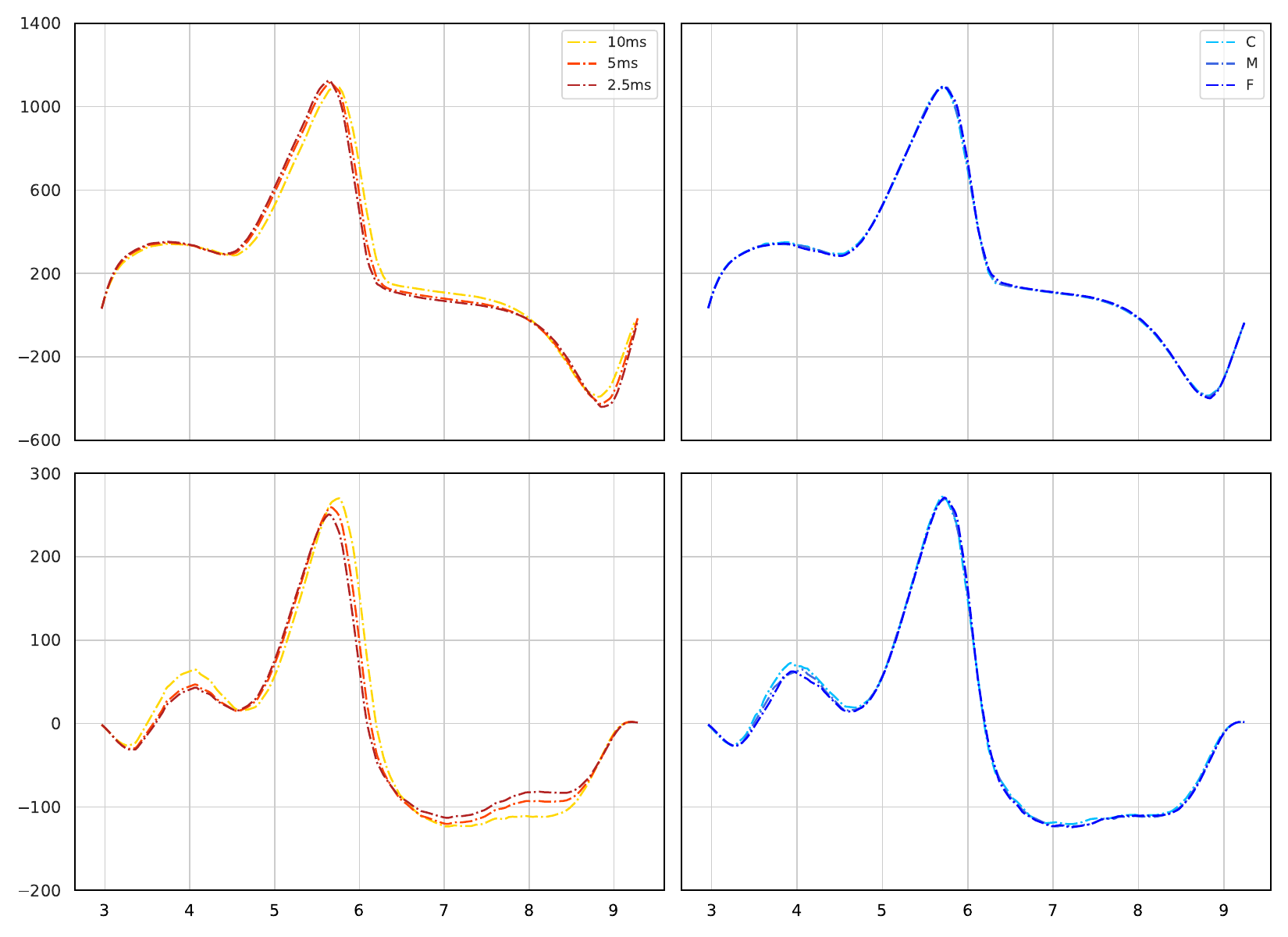}}
    };
    \node[rotate=90] at (-4.1,-1.4) {$\bm{v}_y$ (mm/s)};
    \node[rotate=90] at (-4.1,1.55) {$\bm{v}_x$ (mm/s)};
    \node at (-1.65,-3.1) {y coordinate (mm)};
    \node at (2.15,-3.1) {y coordinate (mm)};
    \end{tikzpicture}
    \caption{Reference systolic velocity profiles ($t=1.08\,s$) along the $y$ axis crossing the origin of the torus (cf. Figure 1 of the main manuscript) for the R2 compliant configuration. The influence of the employed timestep is given on the left, while the impact of the mesh resolution is shown on the right. The medium (M) mesh corresponds to the one used for the reported R2 simulations. The coarse and fine meshes feature $0.7$ / $0.3\,M$ and $1.9$ / $0.9\,M$ fluid / solid elements, respectively.}
    \label{fig:supp_fig1}
\end{figure}


\begin{thebibliography}{10}
\providecommand \doibase [0]{http://dx.doi.org/}%

\bibitem{Ing+00}
Ingall T, Asplund K, Mähönen M, Bonita R. A Multinational Comparison of
  Subarachnoid Hemorrhage Epidemiology in the WHO MONICA Stroke Study. {\it
  Stroke} 2000\string; 31(5)\string: 1054-1061.
\newblock \href {\doibase 10.1161/01.STR.31.5.1054} {doi:
  10.1161/01.STR.31.5.1054}

\bibitem{Vla+11}
Vlak MH, Algra A, Brandenburg R, Rinkel GJ. Prevalence of unruptured
  intracranial aneurysms, with emphasis on sex, age, comorbidity, country, and
  time period: a systematic review and meta-analysis. {\it The Lancet
  Neurology} 2011\string; 10(7)\string: 626-636.
\newblock \href {\doibase https://doi.org/10.1016/S1474-4422(11)70109-0} {doi:
  https://doi.org/10.1016/S1474-4422(11)70109-0}

\bibitem{Kot+12}
Kotowski M, Naggara O, Darsaut T, et al. Safety and occlusion rates of surgical
  treatment of unruptured intracranial aneurysms: A systematic review and
  meta-analysis of the literature from 1990 to 2011. {\it Journal of neurology,
  neurosurgery, and psychiatry} 2012\string; 84.
\newblock \href {\doibase 10.1136/jnnp-2011-302068} {doi:
  10.1136/jnnp-2011-302068}

\bibitem{Nag+12}
Naggara ON, Lecler A, Oppenheim C, Meder JF, Raymond J. Endovascular Treatment
  of Intracranial Unruptured Aneurysms: A Systematic Review of the Literature
  on Safety with Emphasis on Subgroup Analyses. {\it Radiology} 2012\string;
  263(3)\string: 828-835.
\newblock PMID: 22623696\href {\doibase 10.1148/radiol.12112114} {doi:
  10.1148/radiol.12112114}

\bibitem{Nie+18}
Niemann U, Berg P, Niemann A, et al. Rupture Status Classification of
  Intracranial Aneurysms Using Morphological Parameters. {\it In Proceedings of
  the IEEE 31st International Symposium on Computer-Based Medical Systems}
  2018\string: 48-53.
\newblock \href {\doibase 10.1109/CBMS.2018.00016} {doi:
  10.1109/CBMS.2018.00016}

\bibitem{Ceb+14}
Cebral J, Vázquez M, Sforza D, et al. Analysis of hemodynamics and wall
  mechanics at sites of cerebral aneurysm rupture. {\it Journal of
  neurointerventional surgery} 2014\string; 7.
\newblock \href {\doibase 10.1136/neurintsurg-2014-011247} {doi:
  10.1136/neurintsurg-2014-011247}

\bibitem{Men+14}
Meng H, Tutino V, Xiang J, Siddiqui A. High WSS or Low WSS? Complex
  Interactions of Hemodynamics with Intracranial Aneurysm Initiation, Growth,
  and Rupture: Toward a Unifying Hypothesis. {\it American Journal of
  Neuroradiology} 2014\string; 35(7)\string: 1254--1262.
\newblock \href {\doibase 10.3174/ajnr.A3558} {doi: 10.3174/ajnr.A3558}

\bibitem{Jan+15}
Janiga G, Berg P, Sugiyama S, Kono K, Steinman D. The Computational Fluid
  Dynamics Rupture Challenge 2013{\textemdash}Phase I: Prediction of Rupture
  Status in Intracranial Aneurysms. {\it American Journal of Neuroradiology}
  2015\string; 36(3)\string: 530--536.
\newblock \href {\doibase 10.3174/ajnr.A4157} {doi: 10.3174/ajnr.A4157}

\bibitem{Tor+09}
Torii R, Oshima M, Kobayashi T, Takagi K, Tezduyar T. Fluid-structure
  interaction modeling of blood flow and cerebral aneurysm: Significance of
  artery and aneurysm shapes. {\it Computer Methods in Applied Mechanics and
  Engineering} 2009\string; 198\string: 3613-3621.
\newblock \href {\doibase 10.1016/j.cma.2008.08.020} {doi:
  10.1016/j.cma.2008.08.020}

\bibitem{Baz+10}
Bazilevs Y, Hsu MC, Zhang Y, et al. A fully-coupled fluid-structure interaction
  simulation of cerebral aneurysms. {\it Computational Mechanics} 2010\string;
  46\string: 3-16.
\newblock \href {\doibase 10.1007/s00466-009-0421-4} {doi:
  10.1007/s00466-009-0421-4}

\bibitem{Tor+08}
Torii R, Oshima M, Kobayashi T, Takagi K, Tezduyar T. Fluid-structure
  interaction modeling of a patient-specific cerebral aneurysm: influence of
  structural modeling. {\it Computational Mechanics} 2008\string; 43\string:
  151-159.
\newblock \href {\doibase 10.1007/s00466-008-0325-8} {doi:
  10.1007/s00466-008-0325-8}

\bibitem{Tor+10}
Torii R, Oshima M, Kobayashi T, Takagi K, Tezduyar TE. Influence of wall
  thickness on fluid–structure interaction computations of cerebral
  aneurysms. {\it International Journal for Numerical Methods in Biomedical
  Engineering} 2010\string; 26(3-4)\string: 336-347.
\newblock \href {\doibase https://doi.org/10.1002/cnm.1289} {doi:
  https://doi.org/10.1002/cnm.1289}

\bibitem{VoS+16}
Voß S, Saalfeld S, Hoffmann T, et al. Fluid-Structure Simulations of a
  Ruptured Intracranial Aneurysm: Constant versus Patient-Specific Wall
  Thickness. {\it Computational and Mathematical Methods in Medicine}
  2016\string; 2016\string: 1-8.
\newblock \href {\doibase 10.1155/2016/9854539} {doi: 10.1155/2016/9854539}

\bibitem{Val+13}
Valencia A, Burdiles PA, Ignat M, et al. Fluid Structural Analysis of Human
  Cerebral Aneurysm Using Their Own Wall Mechanical Properties. {\it
  Computational and Mathematical Methods in Medicine} 2013\string; 2013.

\bibitem{Van+15}
Vanrossomme A, Eker O, Thiran JP, Courbebaisse G, Zouaoui~Boudjeltia K.
  {Intracranial Aneurysms: Wall Motion Analysis for Prediction of Rupture}.
  {\it {American Journal of Neuroradiology}} 2015\string; 36(10)\string:
  1796-1802.
\newblock \href {\doibase 10.3174/ajnr.A4310} {doi: 10.3174/ajnr.A4310}

\bibitem{Hay+13}
Hayakawa M, Tanaka T, Sadato A, et al. Detection of Pulsation in Unruptured
  Cerebral Aneurysms by ECG-Gated 3D-CT Angiography (4D-CTA) with 320-Row Area
  Detector CT (ADCT) and Follow-up Evaluation Results: Assessment Based on
  Heart Rate at the Time of Scanning. {\it Clinical neuroradiology}
  2013\string; 24.
\newblock \href {\doibase 10.1007/s00062-013-0236-8} {doi:
  10.1007/s00062-013-0236-8}

\bibitem{Zho+22}
Zhou J, Guo Q, Chen Y, et al. Irregular Pulsation of Intracranial Aneurysm
  Detected by Four-Dimensional CT Angiography and Associated With Small
  Aneurysm Rupture: A Single-Center Prospective Analysis. {\it Frontiers in
  Neurology} 2022\string; 13.
\newblock \href {\doibase 10.3389/fneur.2022.809286} {doi:
  10.3389/fneur.2022.809286}

\bibitem{Sta+21}
Stam L, Aquarius R, De~Jong G, Slump C, Meijer F, Boogaarts J. A review on
  imaging techniques and quantitative measurements for dynamic imaging of
  cerebral aneurysm pulsations. {\it Scientific Reports} 2021\string;
  11\string: 2175.
\newblock \href {\doibase 10.1038/s41598-021-81753-z} {doi:
  10.1038/s41598-021-81753-z}

\bibitem{Fin+13}
Finol E, Raut S, Jana A, oliveira dV, Muluk S. The Importance of
  Patient-Specific Regionally Varying Wall Thickness in Abdominal Aortic
  Aneurysm Biomechanics. {\it Journal of biomechanical engineering}
  2013\string; 135.
\newblock \href {\doibase 10.1115/1.4024578} {doi: 10.1115/1.4024578}

\bibitem{Isa+08}
Isaksen J, Bazilevs Y, Kvamsdal T, et al. Determination of Wall Tension in
  Cerebral Artery Aneurysms by Numerical Simulation. {\it Stroke; a journal of
  cerebral circulation} 2008\string; 39\string: 3172-8.
\newblock \href {\doibase 10.1161/STROKEAHA.107.503698} {doi:
  10.1161/STROKEAHA.107.503698}

\bibitem{SO78}
Suzuki J, Ohara H. Clinicopathological study of cerebral aneurysms: Origin,
  rupture, repair, and growth. {\it Journal of Neurosurgery} 1978\string;
  48(4)\string: 505 - 514.
\newblock \href {\doibase 10.3171/jns.1978.48.4.0505} {doi:
  10.3171/jns.1978.48.4.0505}

\bibitem{Kleinloog2018}
Kleinloog R, Zwanenburg J, Schermers B, et al. Quantification of Intracranial
  Aneurysm Volume Pulsation with 7T MRI. {\it American Journal of
  Neuroradiology} 2018\string; 39.
\newblock \href {\doibase 10.3174/ajnr.A5546} {doi: 10.3174/ajnr.A5546}

\bibitem{Illies2016}
Illies T, Säring D, Kinoshita M, et al. Feasibility of Quantification of
  Intracranial Aneurysm Pulsation with 4D CTA with Manual and Computer-Aided
  Post-Processing. {\it PLOS ONE} 2016\string; 11.
\newblock \href {\doibase 10.1371/journal.pone.0166810} {doi:
  10.1371/journal.pone.0166810}

\bibitem{Rob+15}
Robertson AM, Duan X, Aziz KM, Hill MR, Watkins SC, Cebral JR. Diversity in the
  Strength and Structure of Unruptured Cerebral Aneurysms. {\it Annals of
  Biomedical Engineering} 2015\string; 43(7)\string: 1502-1515.

\bibitem{Mer+10}
Baharoglu M, Schirmer C, Hoit D, Gao B, Malek A. Aneurysm Inflow-Angle as a
  Discriminant for Rupture in Sidewall Cerebral Aneurysms Morphometric and
  Computational Fluid Dynamic Analysis. {\it Stroke; a journal of cerebral
  circulation} 2010\string; 41\string: 1423-30.
\newblock \href {\doibase 10.1161/STROKEAHA.109.570770} {doi:
  10.1161/STROKEAHA.109.570770}

\bibitem{Has+05}
Hassan T, Timofeev E, Saito T, et al. A proposed parent vessel geometry-based
  categorization of saccular intracranial aneurysms: Computational flow
  dynamics analysis of the risk factors for lesion rupture. {\it Journal of
  neurosurgery} 2005\string; 103\string: 662-80.
\newblock \href {\doibase 10.3171/jns.2005.103.4.0662} {doi:
  10.3171/jns.2005.103.4.0662}

\bibitem{Ram+13}
Ramalho S, Moura AB, Gambaruto AM, Sequeira A. {\it Influence of Blood Rheology
  and Outflow Boundary Conditions in Numerical Simulations of Cerebral
  Aneurysms}\string: 149--175; New York, NY: Springer New York .
\newblock 2013

\bibitem{formaggia}
Formaggia L, Gerbeau JF, Nobile F, Quarteroni A. {On the Coupling of 3D and 1D
  Navier-Stokes Equations for Flow Problems in Compliant Vessels}.
  2000(RR-3862).
\newblock Projet M3N.

\bibitem{Carotid}
Baz R, Scheau C, Cosmin N, Bordei P. Morphometry of the Entire Internal Carotid
  Artery on CT Angiography. {\it Medicina} 2021\string; 57\string: 832.
\newblock \href {\doibase 10.3390/medicina57080832} {doi:
  10.3390/medicina57080832}

\bibitem{Day}
Day AL. Aneurysms of the ophthalmic segment: A clinical and anatomical
  analysis. {\it Journal of Neurosurgery} 1990\string; 72(5)\string: 677 - 691.
\newblock \href {\doibase 10.3171/jns.1990.72.5.0677} {doi:
  10.3171/jns.1990.72.5.0677}

\bibitem{INTRA}
Yang X, Xia D, Kin T, Igarashi T. IntrA: 3D Intracranial Aneurysm Dataset for
  Deep Learning. In: ; 2020.

\bibitem{Ford2005Inflow}
Ford MD, Alperin N, Lee SH, Holdsworth DW, Steinman DA. Characterization of
  volumetric flow rate waveforms in the normal internal carotid and vertebral
  arteries. {\it Physiological Measurement} 2005\string; 26(4)\string: 477.
\newblock \href {\doibase 10.1088/0967-3334/26/4/013} {doi:
  10.1088/0967-3334/26/4/013}

\bibitem{Ceb+15}
Cebral J, Duan X, Chung B, Putnam C, Aziz K, Robertson A. Wall Mechanical
  Properties and Hemodynamics of Unruptured Intracranial Aneurysms. {\it
  American Journal of Neuroradiology} 2015\string; 36.
\newblock \href {\doibase 10.3174/ajnr.A4358} {doi: 10.3174/ajnr.A4358}

\bibitem{Lau+20}
Laurence D, Homburg H, Yan F, et al. A Pilot Study on Biaxial Mechanical,
  Collagen Microstructural, and Morphological Characterizations of a Resected
  Human Intracranial Aneurysm Tissue. {\it SSRN Electronic Journal} 2020.
\newblock \href {\doibase 10.2139/ssrn.3577288} {doi: 10.2139/ssrn.3577288}

\bibitem{Aco+21}
Acosta JM, Cayron AF, Dupuy N, et al. {Effect of Aneurysm and Patient
  Characteristics on Intracranial Aneurysm Wall Thickness}. {\it Frontiers in
  Cardiovascular Medicine} 2021\string; 8\string: 775307.
\newblock \href {\doibase 10.3389/fcvm.2021.775307} {doi:
  10.3389/fcvm.2021.775307}

\bibitem{HGO2000}
Holzapfel GA, Gasser TC, Ogden RW. {A New Constitutive Framework for Arterial
  Wall Mechanics and a Comparative Study of Material Models}. {\it {Journal of
  elasticity and the physical science of solids}} 2000\string; 61(1)\string:
  1-48.

\bibitem{Robertson2009}
Robertson AM, Sequeira A, Owens RG. {\it Rheological models for blood}\string:
  211--241; Milano: Springer Milan .
\newblock 2009

\bibitem{Fro+04}
Frösen J, Piippo A, Paetau A, et al. Remodeling of Saccular Cerebral Artery
  Aneurysm Wall Is Associated With Rupture Histological Analysis of 24
  Unruptured and 42 Ruptured Cases. {\it Stroke; a journal of cerebral
  circulation} 2004\string; 35\string: 2287-93.
\newblock \href {\doibase 10.1161/01.STR.0000140636.30204.da} {doi:
  10.1161/01.STR.0000140636.30204.da}

\bibitem{Men+07}
Meng H, Wang Z, Hoi Y, et al. Complex Hemodynamics at the Apex of an Arterial
  Bifurcation Induces Vascular Remodeling Resembling Cerebral Aneurysm
  Initiation. {\it Stroke; a journal of cerebral circulation} 2007\string;
  38\string: 1924-31.
\newblock \href {\doibase 10.1161/STROKEAHA.106.481234} {doi:
  10.1161/STROKEAHA.106.481234}

\bibitem{MAl99}
Malek A. Hemodynamic Shear Stress and Its Role in Atherosclerosis. {\it JAMA}
  1999\string; 282\string: 2035.
\newblock \href {\doibase 10.1001/jama.282.21.2035} {doi:
  10.1001/jama.282.21.2035}

\bibitem{gmsh}
Geuzaine C, Remacle JF. Gmsh: A 3-{D} Finite Element Mesh Generator with
  built-in Pre- and Post-Processing Facilities. {\it International Journal for
  Numerical Methods in Engineering} 2009.
\newblock \href {\doibase 10.1002/nme.2579} {doi: 10.1002/nme.2579}

\bibitem{ALEorigin}
Hirt C, Amsden A, Cook J. An arbitrary Lagrangian-Eulerian computing method for
  all flow speeds. {\it Journal of Computational Physics} 1974\string;
  14(3)\string: 227-253.
\newblock \href {\doibase https://doi.org/10.1016/0021-9991(74)90051-5} {doi:
  https://doi.org/10.1016/0021-9991(74)90051-5}

\bibitem{VMS}
Hachem E, Rivaux B, Kloczko T, Digonnet H, Coupez T. Stabilized finite element
  method for incompressible flows with high Reynolds number. {\it Journal of
  Computational Physics} 2010\string; 229\string: 8643-8665.
\newblock \href {\doibase 10.1016/j.jcp.2010.07.030} {doi:
  10.1016/j.jcp.2010.07.030}

\bibitem{babuvska1971error}
Babu{\v{s}}ka I. Error-bounds for finite element method. {\it Numerische
  Mathematik} 1971\string; 16(4)\string: 322--333.

\bibitem{HABCHI2013306}
Habchi C, Russeil S, Bougeard D, et al. Partitioned solver for strongly coupled
  fluid–structure interaction. {\it Computers \& Fluids} 2013\string;
  71\string: 306-319.
\newblock \href {\doibase https://doi.org/10.1016/j.compfluid.2012.11.004}
  {doi: https://doi.org/10.1016/j.compfluid.2012.11.004}

\bibitem{Sha20}
Shamanskiy A, Simeon B. Mesh moving techniques in fluid-structure interaction:
  robustness, accumulated distortion and computational efficiency. {\it
  Computational Mechanics} 2021\string; 67\string: 1-18.
\newblock \href {\doibase 10.1007/s00466-020-01950-x} {doi:
  10.1007/s00466-020-01950-x}

\bibitem{ElasticMeshMove}
Chiandussi G, Bugeda G, Oñate E. A simple method for automatic update of
  finite element meshes. {\it Communications in Numerical Methods in
  Engineering} 2000\string; 16(1)\string: 1-19.
\newblock \href {\doibase https://doi.org/10.1002} {doi:
  https://doi.org/10.1002}

\bibitem{SimoTaylor}
Simo J, Taylor R, Pister K. Variational and projection methods for the volume
  constraint in finite deformation elasto-plasticity. {\it Computer Methods in
  Applied Mechanics and Engineering} 1985\string; 51(1)\string: 177-208.
\newblock \href {\doibase https://doi.org/10.1016/0045-7825(85)90033-7} {doi:
  https://doi.org/10.1016/0045-7825(85)90033-7}

\bibitem{NEMER2021113923}
Nemer R, Larcher A, Coupez T, Hachem E. Stabilized finite element method for
  incompressible solid dynamics using an updated Lagrangian formulation. {\it
  Computer Methods in Applied Mechanics and Engineering} 2021\string;
  384\string: 113923.
\newblock \href {\doibase https://doi.org/10.1016/j.cma.2021.113923} {doi:
  https://doi.org/10.1016/j.cma.2021.113923}

\bibitem{FELIPPA198061}
Felippa C, Park K. Staggered transient analysis procedures for coupled
  mechanical systems: Formulation. {\it Computer Methods in Applied Mechanics
  and Engineering} 1980\string; 24(1)\string: 61-111.
\newblock \href {\doibase https://doi.org/10.1016/0045-7825(80)90040-7} {doi:
  https://doi.org/10.1016/0045-7825(80)90040-7}

\bibitem{CAUSIN20054506}
Causin P, Gerbeau J, Nobile F. Added-mass effect in the design of partitioned
  algorithms for fluid–structure problems. {\it Computer Methods in Applied
  Mechanics and Engineering} 2005\string; 194(42)\string: 4506-4527.
\newblock \href {\doibase https://doi.org/10.1016/j.cma.2004.12.005} {doi:
  https://doi.org/10.1016/j.cma.2004.12.005}

\bibitem{Forster}
Förster C, Wall W, Ramm E. Artificial added mass instabilities in sequential
  staggered coupling of nonlinear structures and incompressible viscous flows.
  {\it Computer Methods in Applied Mechanics and Engineering} 2007\string;
  196\string: 1278-1293.
\newblock \href {\doibase 10.1016/j.cma.2006.09.002} {doi:
  10.1016/j.cma.2006.09.002}

\bibitem{BREUER2012107}
Breuer M, {De Nayer} G, Münsch M, Gallinger T, Wüchner R. Fluid–structure
  interaction using a partitioned semi-implicit predictor–corrector coupling
  scheme for the application of large-eddy simulation. {\it Journal of Fluids
  and Structures} 2012\string; 29\string: 107-130.
\newblock \href {\doibase https://doi.org/10.1016/j.jfluidstructs.2011.09.003}
  {doi: https://doi.org/10.1016/j.jfluidstructs.2011.09.003}

\bibitem{Eken}
Eken A, Sahin M. A parallel monolithic algorithm for the numerical simulation
  of large-scale fluid structure interaction problems. {\it International
  Journal for Numerical Methods in Fluids} 2015\string; 80(12)\string: 687-714.
\newblock \href {\doibase https://doi.org/10.1002/fld.4169} {doi:
  https://doi.org/10.1002/fld.4169}

\bibitem{Kuttler}
Küttler U, Wall W. Fixed-point fluid-structure interaction solvers with
  dynamic relaxation. {\it Computational Mechanics} 2008\string; 43\string:
  61-72.
\newblock \href {\doibase 10.1007/s00466-008-0255-5} {doi:
  10.1007/s00466-008-0255-5}

\bibitem{MALAN2013426}
Malan A, Oxtoby O. An accelerated, fully-coupled, parallel 3D hybrid
  finite-volume fluid–structure interaction scheme. {\it Computer Methods in
  Applied Mechanics and Engineering} 2013\string; 253\string: 426-438.
\newblock \href {\doibase https://doi.org/10.1016/j.cma.2012.09.004} {doi:
  https://doi.org/10.1016/j.cma.2012.09.004}

\bibitem{Ryzha}
Ryzhakov PB, Marti J, Dialami N. A Unified Arbitrary Lagrangian–Eulerian
  Model for Fluid–Structure Interaction Problems Involving Flows in Flexible
  Channels. {\it J. Sci. Comput.} 2022\string; 90(3).
\newblock \href {\doibase 10.1007/s10915-021-01748-w} {doi:
  10.1007/s10915-021-01748-w}

\bibitem{LOZOVSKIY}
Lozovskiy A, Olshanskii MA, Vassilevski YV. Analysis and assessment of a
  monolithic FSI finite element method. {\it Computers \& Fluids} 2019\string;
  179\string: 277-288.
\newblock \href {\doibase https://doi.org/10.1016/j.compfluid.2018.11.004}
  {doi: https://doi.org/10.1016/j.compfluid.2018.11.004}

\bibitem{TurekAnev}
Turek S, Hron J, M{\'a}dl{\'i}k M, Razzaq M, Wobker H, Acker JF. Numerical
  Simulation and Benchmarking of a Monolithic Multigrid Solver for
  Fluid-Structure Interaction Problems with Application to Hemodynamics. In:
  Bungartz HJ, Mehl M, Sch{\"a}fer M. \kern-2pt, eds. {\it Fluid Structure
  Interaction II}Springer Berlin Heidelberg; 2010; Berlin, Heidelberg\string:
  193--220.

\bibitem{HMY06}
Hashimoto T, Meng H, Young W. Intracranial aneurysms: Links among inflammation,
  hemodynamics and vascular remodeling. {\it Neurological research}
  2006\string; 28\string: 372-80.
\newblock \href {\doibase 10.1179/016164106X14973} {doi:
  10.1179/016164106X14973}

\bibitem{Gut+19}
Grande~Gutierrez N, Mathew M, McCrindle B, et al. Hemodynamic variables in
  aneurysms are associated with thrombotic risk in children with Kawasaki
  disease. {\it International Journal of Cardiology} 2019\string; 281.
\newblock \href {\doibase 10.1016/j.ijcard.2019.01.092} {doi:
  10.1016/j.ijcard.2019.01.092}

\bibitem{Bep+20}
Beppu M, Tsuji M, Ishida F, Shirakawa M, Suzuki H, Yoshimura S. Computational
  Fluid Dynamics Using a Porous Media Setting Predicts Outcome after
  Flow-Diverter Treatment. {\it AJNR. American journal of neuroradiology}
  2020\string; 41.
\newblock \href {\doibase 10.3174/ajnr.A6766} {doi: 10.3174/ajnr.A6766}

\bibitem{Par+23}
Paritala P, Anbananthan H, Hautaniemi J, et al. Reproducibility of the
  computational fluid dynamic analysis of a cerebral aneurysm monitored over a
  decade. {\it Scientific Reports} 2023\string; 13.
\newblock \href {\doibase 10.1038/s41598-022-27354-w} {doi:
  10.1038/s41598-022-27354-w}

\end{thebibliography}
\end{document}